\documentclass[11pt]{article}

\usepackage{arxiv}
\usepackage{pdfpages} % for insert pdfs and graphics

\usepackage{hyperref}       % hyperlinks
\usepackage{url}            % simple URL typesetting
\usepackage{xurl}		   % line breaks
\usepackage{booktabs}       % professional-quality tables
\usepackage{amsfonts}       % blackboard math symbols
\usepackage{nicefrac}       % compact symbols for 1/2, etc.
\usepackage{microtype}      % microtypography
\usepackage{lipsum}
\usepackage{amsmath,amssymb}
\usepackage{setspace}
\usepackage{natbib}
\usepackage{cleveref} 
\usepackage[latin1]{inputenc} % to solve a strange issue in this paper's compiling, not necessarily useful in other files
\usepackage{lscape}

\usepackage{mathpazo}
\usepackage{appendix}
\usepackage{etoolbox}
\BeforeBeginEnvironment{appendices}{\clearpage}

\hypersetup{colorlinks=true,citecolor=blue,linkcolor=blue,anchorcolor =blue,urlcolor=blue}

\title{Google and China's Trade\thanks{Hu: Central University of Finance and Economics (\url{hucui@cufe.edu.cn}). Li: University of Massachusetts (\url{benli36@gmail.com}).} }
\author{
  Cui Hu
   \And
 Ben G. Li
}

\begin{document}
\maketitle

\begin{abstract}
Although Google is blocked in China, Chinese provinces export significantly more to foreign countries that recently searched for them (up to 12 months prior). This attention premium is found mainly at the extensive margin of exports, larger in products that are relatively homogeneous, substitutable, and upstream in the production process, and more pronounced during the COVID pandemic and during the holiday season. The attention premium is not found for Chinese imports from the rest of the world. Our findings attest to online attention as a scarce resource in international trade allocated by importers.  (\textit{JEL Codes: F14, D83})
\end{abstract}

%Having attention of potential buyers is a compelling business advantage. Using Google search frequency to measure attention, we find that Chinese provinces receiving more searches from a foreign country in the preceding months (up to 12 months) export more to the foreign country. This online search premium is found mainly at the extensive margin of exports, larger in products that are homogeneous, substitutable, and upstream in the production process, and more pronounced during the COVID pandemic. Treating the attention received in online search as either a predictor or a driver of trade, our findings provide the first piece of direct, reduced-form evidence of large-scale search activities in the practice of international trade. 

% keywords can be removed

%\newpage

\bigskip
\onehalfspacing

\section{Introduction}

Attention, as a scarce resource in human societies \citep{Kahneman73}, plays a role in international trade.  Countries that host the Olympic Games or the World Cup  see an increase in trade with other nations after the events \citep{RS11, AU14}. Similarly, greater exposure through foreign mass media \citep{CF21} and social media \citep{BGHKRS21} is linked to higher trade volumes. Beyond major events and media exposure, attention is foundational to trade as a whole---buyers and sellers must first be aware of each other before transactions can occur. In this way, attention serves as the starting point for trade. Despite its importance, the role of attention in trade remains underexamined due to the challenges in measuring it.

In this paper, we use the frequency of Google searches to measure attention and examine its role in international trade. Our laboratory is China, where local access to Google is blocked, such that the attention received by China's subnational regions in Google searches comes exclusively from the foreign world. We examine how monthly provincial exports relate to the frequency of the province being searched via Google in foreign countries during preceding, concurrent, and succeeding months. We find that Chinese provinces export significantly more to foreign countries that searched for them in the past 12 months. The search elasticity of trade is 0.85 to 0.88. That is, provinces receiving ten percent more Google searches from a foreign country exported nearly nine percent more to that country. Google searches conducted in the succeeding months reveal no trade relevance. In contrast to exports, imports by Chinese provinces do not correlate with Google searches.

Consider two distinct ways to interpret the above attention premium in trade. (i) Importers, or the fellow citizens they serve, conducted web searches for those provinces \textit{after} developing an interest in their exports. In this case, web searches are  an operational step in trade, predicting trade without directly creating it. (ii) Importers or their fellow citizens developed interest in those provinces \textit{through} web searches. In this case, web searches are the source of the interest, directly creating the trade. Regardless of which interpretation applies, web searches matter in trade because attention is scarce. Importers are not aware of all foreign exporters---and vice versa---so they allocate attention selectively. Whether web searches serve as a tool for executing preexisting attention or as a force that shapes new attention, they reflect how attention is allocated.

Our identification stems from the varying time-series trajectories of Google searches across different province-country pairs. While Google searches and export volume influence each other, we find that export volume is correlated with preceding- and concurrent-month searches, whereas succeeding-month searches show no correlation with export volume. This one-directional correlation is not spuriously driven by confounding factors that affect both export volume and Google searches. Our results cannot be explained by firms strategically marketing their products on search engines. Although firms may advertise their names, brands, or products on search engines, those advertisements do not affect the Google searches for their provinces as keywords. Firms have little incentive to promote their provinces on Google, and governments have even less since they have blocked it.

%Google search data for Chinese provinces can be attributed to specific countries, allowing for the creation of province-country pairs that can be directly integrated with province-country trade data. 

We also find that the attention premium in trade is more nuanced than commonly expected. First, it is primarily evident at the extensive margin. In the trade literature, the expansion of trade through the introduction of new traded varieties, as opposed to the increased traded value of existing varieties, is referred to as extensive margin growth, while the latter is known as intensive margin growth. Our analysis reveals that the export growth following Google searches is driven by the extensive margin rather than the intensive margin.

Second, the attention premium in trade is more pronounced for products that are relatively homogeneous, substitutable, and upstream in the production process. Since switching suppliers for these products incurs lower costs than for others, increased attention to Chinese provinces is more likely to create new trade opportunities for them.

Third, product prices do not respond to Google searches in preceding, concurrent, or succeeding months.\footnote{There is an industrial organization literature that estimates consumers' search costs using price data collected from e-commerce platforms \citep{HS06, AVZ09, HMD09, DHW12, Koulayev14, DelosSantos18, DELS18, JT19}. Our provincial trade data, which are available only by HS product category, do not directly reflect firm-level pricing. We examine the influence of Google searches on HS8-level price dispersion within province-destination-HS4 product groups in \Cref{sec:extensionA}.} Fourth, among export forecasting models, search frequency with a one-month lag is the most effective predictor of exports. Search frequency in later months, even if hypothetically available at the time of forecasting, does not improve forecast accuracy. Fifth, the attention premium in trade grew stronger during the COVID pandemic, likely due to quarantine policies restricting other information channels, especially in-person interactions. Sixth, the attention premium in trade is weaker in the summer and fall compared to winter, which can be attributed to increased holiday-related consumption of Chinese products during the winter season.

Frequencies in web searches have been widely used in the finance literature as a direct measure of attention \citep{MWZ10,DEG11,Tetlock11,VM12,Vozlyublennaia14,AH15,BDI17,KMS17}. Financial markets and international trade markets have the largest trade volume in the real economy. The world markets of goods, which do not have centralized electronic exchanges like the financial markets, have long been known to be full of information frictions.\footnote{Extensive research has demonstrated that information frictions in international trade can be mitigated by advancements in communication technologies. Notable studies include \citet{FW04}, \citet{FMN05}, \citet{PR05}, \citet{BG06}, \citet{CW06}, \citet{Jensen07}, \citet{Aker10}, \citet{LOSV16}, \citet{Fort17}, \citet{JS18}, and \citet{Steinwender18}.} Due to these information frictions, attention allocation is expected to have a greater impact on the trade of goods than on the trade of financial assets. Since attention is unobservable, researchers have studied related behaviors such as price searching \citep{Allen14,Chaney18,EEJKT21,EJTX22}, product marketing \citep{Arkolakis10,FHY24}, traders' networking \citep{RT02}, and peer learning \citep{FT14}. These models offer insights about real-world trade by explaining empirical irregularities unexplained by traditional trade models. However, attention itself remains difficult to observe.  

Two studies directly assess the role of attention in international trade but take a different approach from ours. \citet{DM18} apply the concept of rational inattention in macroeconomics to model and calibrate how importers distribute their attention among exporters. Since gathering information on all potential exporters is impractical, importers in their model begin with limited initial information on observable trade costs and then update their knowledge in a Bayesian manner. \citet{CHL20} present a rare reduced-form finding that indicates the scarcity of attention in international trade. They find that firms with alphabetically earlier names exported more to countries whose languages are linguistically closer to the Latin alphabet. Since these firms appear earlier when importers search trade catalogs and commodity databases, the pattern can be attributed to importers' limited attention span, as they evidently stop searching after selecting the earlier-listed firms. If importers had unlimited attention to search all potential exporters, the alphabetical order of firm names would have no impact on trade patterns.

There is a vast body of literature utilizing Google searches to analyze and predict various offline economic and societal phenomena, including epidemics and pandemics \citep{GMPBSB09, BNWS20, Lampos21}, product and service sales \citep{GHLPW10, CV12}, international migration \citep{BGS20}, consumer panic \citep{KN21}, employment trends \citep{BS22}, economic uncertainty \citep{BBD16}, local corruption \citep{SS13}, and racial animus \citep{Stephens-Davidowitz14}.  Our paper contributes to this literature by expanding the use of Google search data to international trade. International trade is not only the world's most valuable economic activity but also one of the most significant societal phenomena that has transformed human life over the past few centuries.

The rest of the paper is organized as follows. In \Cref{sec:backgrounddata}, we describe the background and sources of our data.  In \Cref{sec:baselinefindings}, we present our baseline specifications and findings. In \Cref{sec:extensions}, we extend our baseline specifications to explore different aspects of our data, highlighting four distinct applications of Google search data in international trade. In \Cref{sec:conclude}, we conclude.

%World Expos

\section{Background and Data \label{sec:backgrounddata}}

\subsection{Google Trend Index (GTI)}

Google is the world's leading search engine. It maintains an estimated 90 percent share of the global web search market, far surpassing competitors such as Bing, Yahoo, and Yandex.\footnote{Market share data are sourced from \citet{Similarweb23}, \citet{Statcounter23}, and \citet{Webb24}.} Google's search engine operates by crawling webpages, collecting text, images, metadata, and links, and storing this information on its servers. The data are then indexed, ranked by its algorithm, and presented to users based on their search queries. Through patented technologies at each stage of this process, Google has retained its position as the dominant search engine worldwide since the 2000s.

In Mainland China, Google has been inaccessible for over a decade and is still unavailable today. The company launched its services there in 2006 but soon faced conflicts with Chinese authorities over politically sensitive search results. In 2010, Google opted to operate only a Hong Kong version within China's jurisdiction. Between 2010 and 2013, Google's products were gradually blocked throughout Mainland China.\footnote{For details of Google leaving Mainland China and the aftermath, see \citet{GP10}, \citet{Sheehan18}, and \citet{Moreno19}. Google considered returning to Mainland China in 2018 by providing censored search results, a project that was terminated by 2019. One can verify the (no) access status of Google in China in real time at \url{https://www.comparitech.com/privacy-security-tools/blockedinchina/google/} (accessed March 16, 2025).} In our context, the inaccessibility of Google in Mainland China creates an identification advantage since the Google searches for Chinese provinces could be exclusively attributed to users in the rest of the world.\footnote{Internet users in Mainland China may use virtual private networks (VPNs) to access Google, though this practice is illegal and subject to intermittent blocking by Chinese internet service providers. The search volume generated through VPNs is minimal and unlikely to impact our province-country level search and trade data analysis.} China's major search engine, Baidu, held a market share of just 0.81 to 0.87 percent \citep{Statcounter23,Webb24} and was seldom used in the rest of the world  due to its obscurity outside China.

Google Trends is a tool by Google that analyzes and tracks the popularity of search keywords over time across different countries. It has been widely used in economic and financial research \citep[e.g.,][]{DEG11,VM12,Vozlyublennaia14,AH15,BDI17}. The tool works as follows. Users input a keyword, a country (or region), and a time period of interest, and then Google Trends returns a frequency index. The frequency index, which we refer to as the Google Trend Index (GTI), represents the keyword's search frequency in the specified country and period. GTI is based on search volumes but accounts for variations in internet usage across countries. Since countries differ in the number of internet users and the quality of internet infrastructure, raw search volumes are not directly comparable. Instead, Google Trends calculates the proportion of searches for a given keyword relative to the total search volume in the specified country. These share values are normalized into GTI for the specified period: zero or little volume is set to 0, the highest to 100, and all other values are scaled proportionally within this range. Google Trends does not disclose raw search volumes for keywords or total search volumes of countries---only the GTI is provided.

GTI operates on a broad match algorithm. For instance, searches for \textit{used automobiles} are factored into the GTI of the keyword \textit{automobile} \citep{CV12}. Similarly, in our context, searches for \textit{Beijing Winter Olympics} and \textit{Beijing government} both contribute to the GTI of the keyword \textit{Beijing}.

The search keywords in our context are Chinese provinces. Mainland China consists of 31 province-level administrative divisions, including 22 provinces, four municipalities (such as Beijing and Shanghai), and five autonomous regions. Henceforth, we refer to all these divisions as provinces. Our data span the period January 2014 to October 2022 on a monthly basis.\footnote{Released in November 2022, ChatGPT, a generative artificial intelligence tool, may have since influenced the search behavior of internet users.} The original GTI data can be considered bilateral, with variation in both the keyword and the searchers' region. Keywords are Chinese provinces, while searchers' regions correspond to foreign countries, producing bilateral Google search data that can be merged with bilateral trade data.

Collecting GTI data for province-product combinations, such as \textit{Beijing machine parts}, may seem appealing. However, due to limitations in Google Trends, this approach is not feasible. Google Trends uses a sampling method to analyze search data and reports GTI only for keywords with sufficient search volume. For province-product combinations, the search volume is typically too low to produce a meaningful GTI value.\footnote{When using the Google Trends website interactively, the error message reads: ``Hmm, your search doesn't have enough data to show here. Please make sure everything is spelled correctly, or try a more general term.'' } \Cref{fig:provinceproduct} illustrates this issue with four examples: searches from the United States for \textit{Beijing} and \textit{Shanghai} combined with either \textit{machine parts} or \textit{cereals}. The GTI data returned by Google Trends consist of either isolated spikes or blank outputs.\footnote{\citet{BGS20} encounter similar challenges in their GTI-based study on international migration. \citet{GKT19} review such challenges in Section 2.5 of their survey on the applications of textual data in economics. } It is worth noting that Beijing and Shanghai are among China's largest provincial economies, and the United States is both the largest user of Google and its home base. Furthermore, machine parts and cereals are generic terms, far more searched than standard product classifications such as ``machinery and mechanical appliances; parts thereof'' (HS Code 84). Additionally, the keyword ``cereals'' encompasses daily consumption items, many of which are unrelated to exports. We intentionally selected these combinations to maximize research volume. However, despite these efforts, no meaningful GTI data was available for these combinations.

\subsection{Patterns in the original GTI data}

\begin{figure}[h!]
%\vspace{30pt}
\caption{Examples of the Original GTI Data} \label{fig:examples}
%\vspace{-20pt}
\hspace*{.5cm}\includegraphics[scale=.75, page=1]{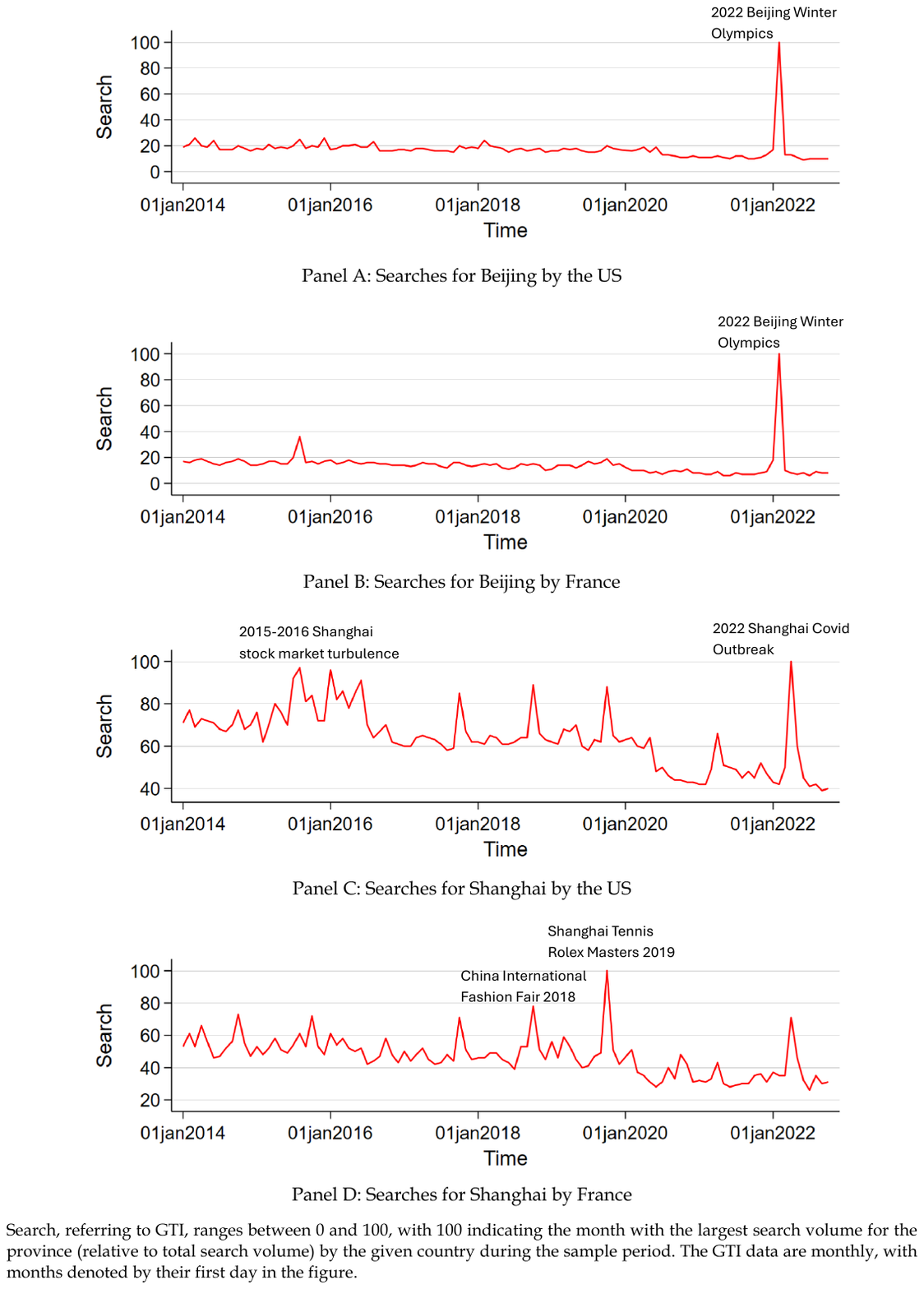}
\vspace{-40pt}
\end{figure}

To illustrate the structure of the GTI data, we present several examples in \Cref{fig:examples}. Panel A of the figure displays the GTI for the keyword \textit{Beijing} searched in the United States over the sample period. The search frequency peaked in February 2022, resulting in a GTI value of 100 for that month. In Panel B, the GTI pertains to the searches for \textit{Beijing} conducted in France over the same period. Although both panels exhibit similar time patterns, the two GTI series are based on distinct ``Beijing shares.'' Specifically, when Google Trends calculates GTI, the search volume for \textit{Beijing} in the United States (France) is divided by the contemporary total search volume of the United States (France). The similarity between the two panels suggests that the Beijing Winter Olympics in February 2022 attracted exceptional attention in searches from both countries, such that the differences in their searches for other terms were not enough to affect the overall time pattern.

Compared to searches for \textit{Beijing}, searches for \textit{Shanghai} exhibit more distinct time patterns between the two countries, as shown in Panels C and D. This suggests that Google users in the United States and France were influenced by different interests. As the country with the largest financial sector and the highest number of reported COVID-related deaths, the United States showed greater interest in Shanghai's stock market fluctuations and COVID outbreak. This interest may also have been driven by geopolitical factors, as populist politicians in the United States heavily criticized China for its economic policies and pandemic responses. In contrast, France demonstrated greater interest in events like the international fashion fair in Shanghai and the Shanghai Tennis Rolex Masters tournament. Shanghai is generally perceived as more commercial, cosmopolitan, and liberal, whereas Beijing is seen as more traditional and politically focused. These differing perceptions are evident in the distinct search patterns observed between the two countries.

\Cref{fig:examples} illustrates both the variations important and unimportant to our analysis. Foreign countries exhibit differing levels of interest in Chinese provinces over time for diverse reasons, which are captured by the GTI and used by us to build an empirical strategy. For example, as the attention paid by the United States to Shanghai rises, peaks, and falls, the attention paid by the United States to Beijing, by France to Shanghai, or by France to Beijing, has distinct fluctuations. These variations in attention across provinces (keywords) and countries (searchers) are linked by us to Shanghai's and Beijing's exports to those countries, serving as the major variations used in our identification. A given province may attract simultaneously elevated attention from multiple countries (e.g., Beijing from the United States and France in February 2022). Such simultaneous changes will be absorbed by province-month fixed effects in our econometric analysis, not serving as the major variations used by our identification. We will elaborate on our empirical strategy in \Cref{sec:spec}.

\subsection{Patterns in the aggregated GTI data}

\begin{figure}[h!]
%\vspace{30pt}
\caption{Chinese Provinces vs. China in Google Searches} \label{fig:provsVSchina}
%\vspace{-10pt}
\hspace*{.5cm}\includegraphics[scale=.75, page=2]{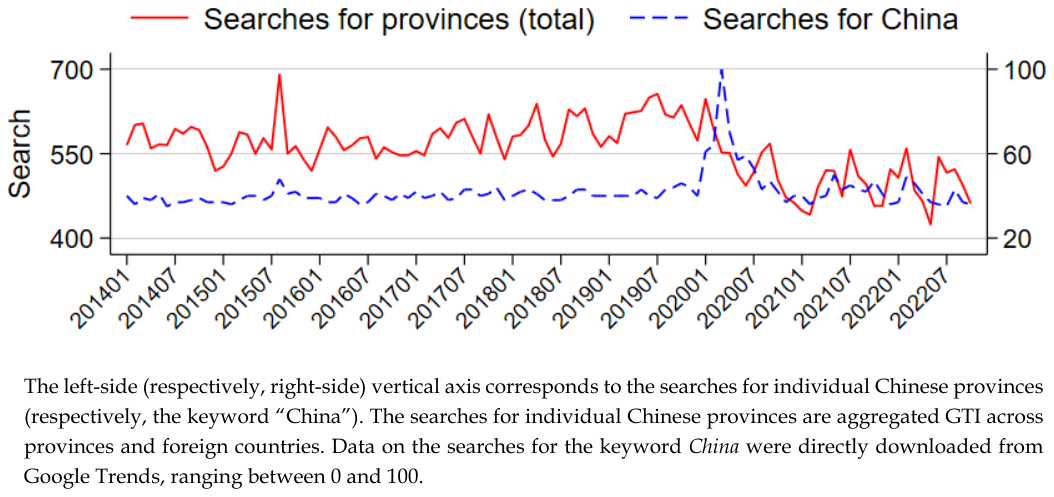}
\vspace{-440pt}
\end{figure}

The original GTI data, when aggregated across provinces and foreign countries, offer insights into the online attention China receives from the rest of the world. First, the searches for Chinese provinces should not be equated with the searches for China as a whole. We downloaded the GTI data for the keyword \textit{China} and plot it over time as the dashed curve in \Cref{fig:provsVSchina}. As shown, the Google searches for China by the rest of the world has remained stable, with an average of 42 out of 100. The solid curve in the figure represents the total search frequency, defined as $TotalProvs_{t} \equiv \sum_j \sum_c GTI_{jct}$. Here, $j$, $c$, and $t$ represent province, country, and month, respectively. Recall that a province-country pair in a given month has a maximum Google search frequency of 100, as measured by GTI. We define GTI$=$100 as a unit called a maximum index month (MIM), meaning that a country searches for a specific province at the maximum frequency in the month. The mean of $TotalProvs_{t}$ is 558.5 MIM. In other words, in an average month during our sample period, the total attention received by Chinese provinces is equivalent to 558.5 province-country pairs reaching the full frequency. The theoretical maximum of $TotalProvs_{t}$ is 5,394 MIM, because the total number of province-country pairs is 31 provinces $\times$ 174 countries = 5,394. That is, on average, a province-country pair maintained approximately one-tenth of its maximum frequency.

The search frequency for the keyword \textit{China} had a spike in early 2020, coinciding with the initial outbreak of the COVID pandemic in the country. In contrast, searches for individual Chinese provinces declined following the outbreak, likely due to China's stringent quarantine policies, which significantly reduced business activities, cultural exchanges, and international travels between China and the rest of the world. While China as a whole continued to capture global attention, there was a decline in foreign engagement involving specific Chinese regions, as reflected in searches related to provincial names. We will further explore the role of Google searches in the pandemic context in \Cref{sec:extensionC}. For now, it is evident that searches for China and its provinces exhibit frequency patterns distinct from each other.

\begin{figure}[h!]
%\vspace{30pt}
\caption{Total Searches for Specific Provinces Conducted by the Foreign World} \label{fig:exampleprovs}
%\vspace{-20pt}
\hspace*{.5cm}\includegraphics[scale=.75, page=3]{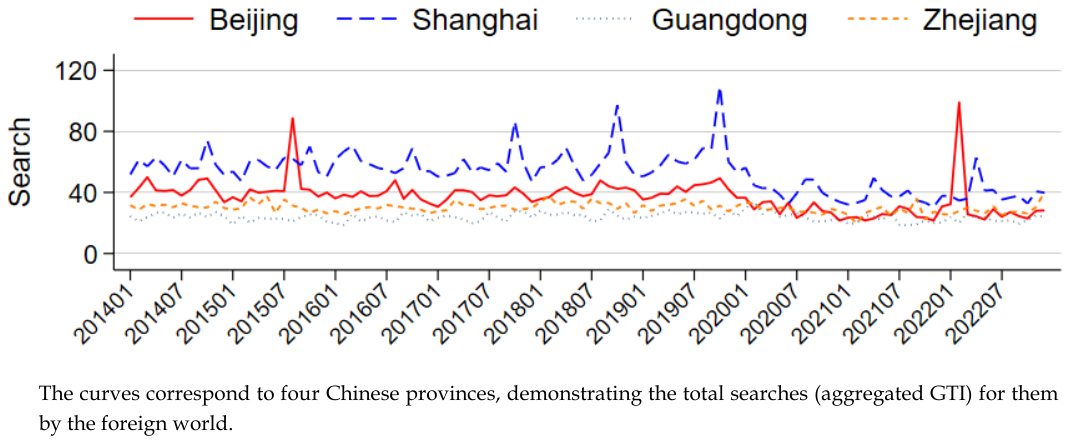}
\vspace{-440pt}
\end{figure}

The $TotalProvs_{t}$, when disaggregated across provinces, represents the total search frequency for each province by the foreign world, expressed as $TotalProvs_{jt} \equiv \sum_c GTI_{jct}$. The average value of $TotalProvs_{jt}$ is 30.9 MIM. The theoretical maximum of $TotalProvs_{jt}$ is 174 MIM (i.e., 174 countries searching for province $j$ in month $t$ at the maximum frequency). In \Cref{fig:exampleprovs}, we illustrate $TotalProvs_{jt}$ for Beijing, Shanghai, and the two largest exporting provinces in China (Guangdong and Zhejiang). As depicted, Shanghai consistently garnered the most attention, followed by Beijing. The additional attention Beijing received compared to the two major exporting provinces was modest, except for a brief surge during the Beijing Winter Olympics.

\begin{figure}[h!]
%\vspace{30pt}
\caption{Total Searches for Chinese Provinces Conducted by Specific Foreign Countries} \label{fig:examplecountries}
%\vspace{-20pt}
\hspace*{.5cm}\includegraphics[scale=.75, page=4]{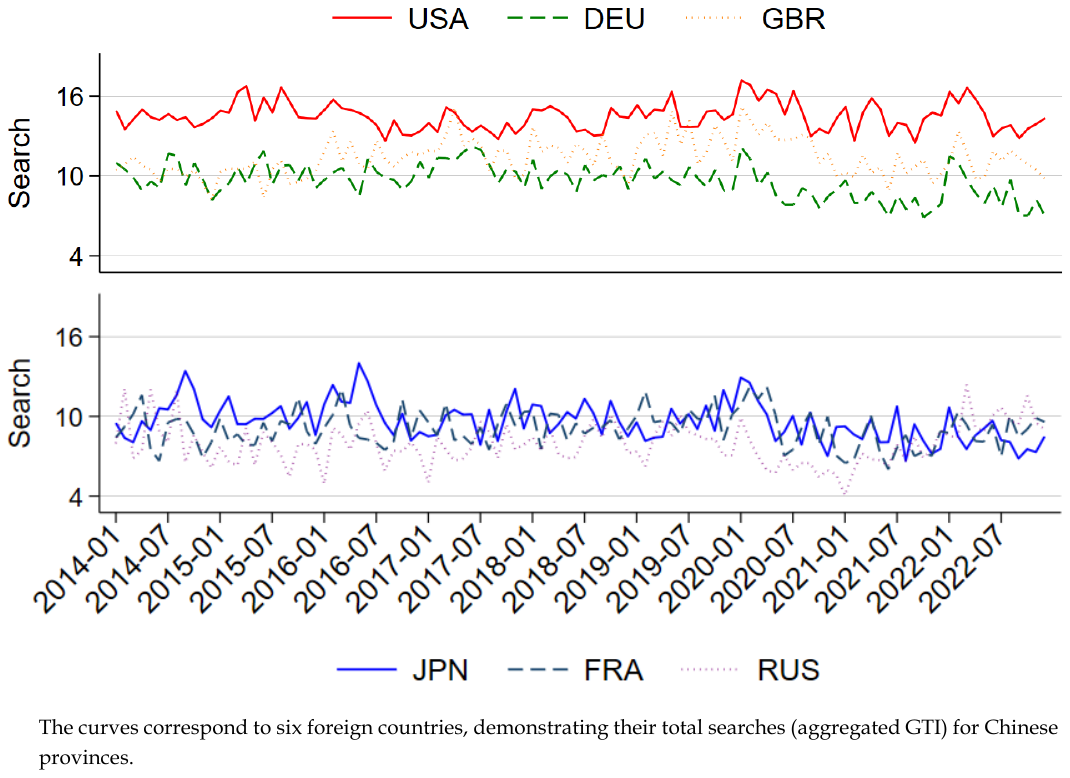}
\vspace{-350pt}
\end{figure}

By disaggregating $TotalProvs_{t}$ across foreign countries, we derive the total search frequency for Chinese provinces conducted by each foreign country, represented as $TotalProvs_{ct} \equiv \sum_j GTI_{jct}$. The average value of $TotalProvs_{ct}$ is 10.4 MIM. The theoretical maximum of $TotalProvs_{ct}$ is 31 MIM (i.e., country $c$ searching for all 31 provinces in month $t$ at the maximum frequency). In \Cref{fig:examplecountries}, we present $TotalProvs_{ct}$ for six of China's major trade partners. Among these, the United States exhibits the highest search frequency, while the remaining countries are similar to each other. As explained earlier, the higher search frequency observed in the United States is not due to its large number of internet users or advanced internet infrastructure---these factors have been normalized away and thus do not influence the GTI data. Instead, it reflects a sustained high search frequency over multiple months relative to its peak-frequency month for the corresponding Chinese province.
 
\subsection{GTI and Trade}
To study the relationship between GTI and trade, we now revert to the original GTI data. We merge the original GTI data, which are at the province-country-month level, with export data from China Customs. \Cref{tab:summarystats} presents descriptive statistics of our working dataset. The export data cover the exports by every province. The original export data, available at the province-country-product-month level (each product referring to a HS4 code), are also used in certain parts of this study. There are 106 months between January 2014 and October 2022, though customs data of January 2020 (the time of China's COVID pandemic outbreak) were counted by the customs agency into those from February 2020 due to logistical chaos resulting from the pandemic. This is why there are only 105 months in our working dataset. We convert the original 0-to-100 range of GTI to 0-to-1 to ease the interpretation of our regression results. 

\begin{table}[h!]
\vspace{20pt}
\caption{Descriptive Statistics} \label{tab:summarystats}
\vspace{-40pt}
\hspace*{0cm}\includegraphics[scale=1, page=1]{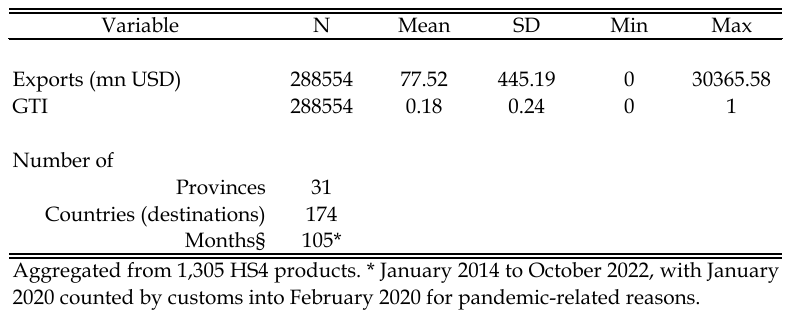}
\vspace{-600pt}
\end{table}

The GTI and export data are both stationary. A panel-data unit root test takes the form of testing the null hypothesis $\phi_p = 0$ in the following equation:
\begin{equation}
\Delta y_{pt} = \phi_p y_{p,t-1} + \gamma_p + \epsilon_{pt}, \label{eq:unitroot}
\end{equation}
where $y_{\bullet}$ is the variable of interest, $p$ indexes the panels, and $t$  indexes time periods. In our context, $y$ is either GTI or logged exports, $p$ is province ($j$), country ($c$), or province-country pair ($jc$), and $t$ is month. \citet*{IPS03} developed a test (henceforth, the IPS test) that accommodates unbalanced panels, making it suitable for our context, as exporting provinces have varying sets of destination countries. The IPS test assumes a fixed number of time periods, with the panel count either remaining constant or approaching infinity, making it well-suited to the structure of trade statistics. Additionally, the IPS test includes variants that allow for panel-specific linear trends and additional lags of the dependent variable. We apply the IPS test to the original (bilateral) GTI and export data and find that unit roots are rejected at all conventional significance levels. The results remain consistent when we incorporate time trends or additional lags of GTI. Furthermore, when the GTI and export data are aggregated across either provinces or countries to create unilateral datasets (as in the previous subsection), unit roots are still not detected. Detailed test results are reported in \Cref{tab:unitroot}.

\section{Baseline Findings\label{sec:baselinefindings}}

\subsection{Specification \label{sec:spec}}

Our empirical strategy is to use current, lagged, and forwarded Google searches for Chinese provinces conducted in foreign countries to explain provincial export volume. We analyze how Google searches over various time frames---before and after the time of the trade data---correlate with export volume. Let \( X_{jct} \) represent the export volume from province \( j \) to country \( c \) in month \( t \), and \( GTI_{jc\tau} \) denote the aforementioned frequency of Google searches (GTI) conducted in country \( c \) for province \( j \) in month \( \tau \). Our primary specification is
\begin{equation}
\ln X_{jct} = \sum_{\tau = t-12}^{t+12} \beta_\tau GTI_{jc\tau} + \lambda_{jc} + \lambda_{jt} + \lambda_{ct} + \epsilon_{jct}, \label{eq:reg}
\end{equation}
where $\lambda_{jc}$, $\lambda_{jt}$, and $\lambda_{ct}$ are province-country, province-month, and country-month fixed effects, respectively, and $\epsilon_{jct}$ is a classic error term. To mitigate collinearity, we include $t$ at intervals of three months, spanning up to $\pm 12$ months, instead of incorporating every month within the range.

Our empirical strategy builds on the varying time-series trajectories of GTI across different province-country pairs. Each country has fluctuating interests in different Chinese provinces over time. When a country's attention to a province rises, the attention from other countries to other provinces does not simultaneously change in the same manner. For each month in the export data, we examine the GTI values of the preceding, concurrent, and succeeding months. By comparing these differing time-series trajectories across province-country pairs, we can identify the statistical association between GTI and exports within specific time windows. This time-window approach resembles an event study setting. That is, the search frequencies for different province-country pairs peak at different times, and specification \eqref{eq:reg} exploits these peak events and the corresponding pre- and post-event time windows.

Google searches, as a measure of attention, were directed toward Chinese provinces by foreign countries due to the interest of their importers (or the fellow citizens they serve) in the provinces' exports. This interest could either predate the searches or emerge at the time the searches were conducted. When interest preexists, searches are part of the trading process that facilitates transactions. When interest arises anew, searches act as a demand shock for Chinese exports. The difference between these two scenarios is not crucial for our study, as our objective is to assess the relevance of web searches to trade. In practice, the interest underlying web searches is typically a combination of both scenarios. Regardless, $\beta_{\tau}$, if found statistically different from zero, would reveal a statistical association between Google searches and export volume. Thus, differentiating between the two scenarios is unnecessary before establishing the statistical relationship.

It is important to note that the keywords in Google search results are the names of provinces, not the names of firms or their product brands. For example, importers searching for \textit{Lenovo} are unlikely to use \textit{Beijing} as a keyword, even though it is Lenovo's headquarters, just as those looking for \textit{Huawei} are unlikely to search for \textit{Guangdong}, where Huawei is based. Exporters have no incentive to promote their provinces, nor do provincial governments---especially considering that they have blocked Google. Using provinces as search keywords mitigates potential confounding factors. Advertisements on search engines would be a key confounding factor if the analysis were conducted at the firm level using Google search results specific to firm-level keywords. 

The inclusion of multiple GTI variables as regressors also helps mitigate confounding factors. Following \citet{CV12}, we include GTI up to $\pm 12$ months. This joint inclusion leverages the residual variation among these variables, allowing us to observe their differing relevance to trade. We include forwarded GTI variables as placebos. As demonstrated later, the insignificance of forwarded GTI, in contrast to the significance of lagged GTI, strengthens our confidence that the findings are not driven by persistence in web-search behaviors.

The fixed effects in equation \eqref{eq:reg} account for bilateral time-invariant characteristics (e.g., province-country distance) and unilateral time-variant factors (e.g., population and GDP). Furthermore, their combined application effectively demeans GTI along these dimensions. Specifically, provinces frequently searched by foreigners are compared to their monthly means, countries with sustained interest in Chinese provinces are compared against their monthly means, and province-country pairs with strong search connections are compared relative to their pairwise means.

\subsection{Main results}

\begin{figure}[h!]
%\vspace{30pt}
\caption{Google Searches and Export Volume: Basic Patterns} \label{fig:correlation}
%\vspace{-20pt}
\hspace*{-0.5cm}\includegraphics[scale=.85, page=5]{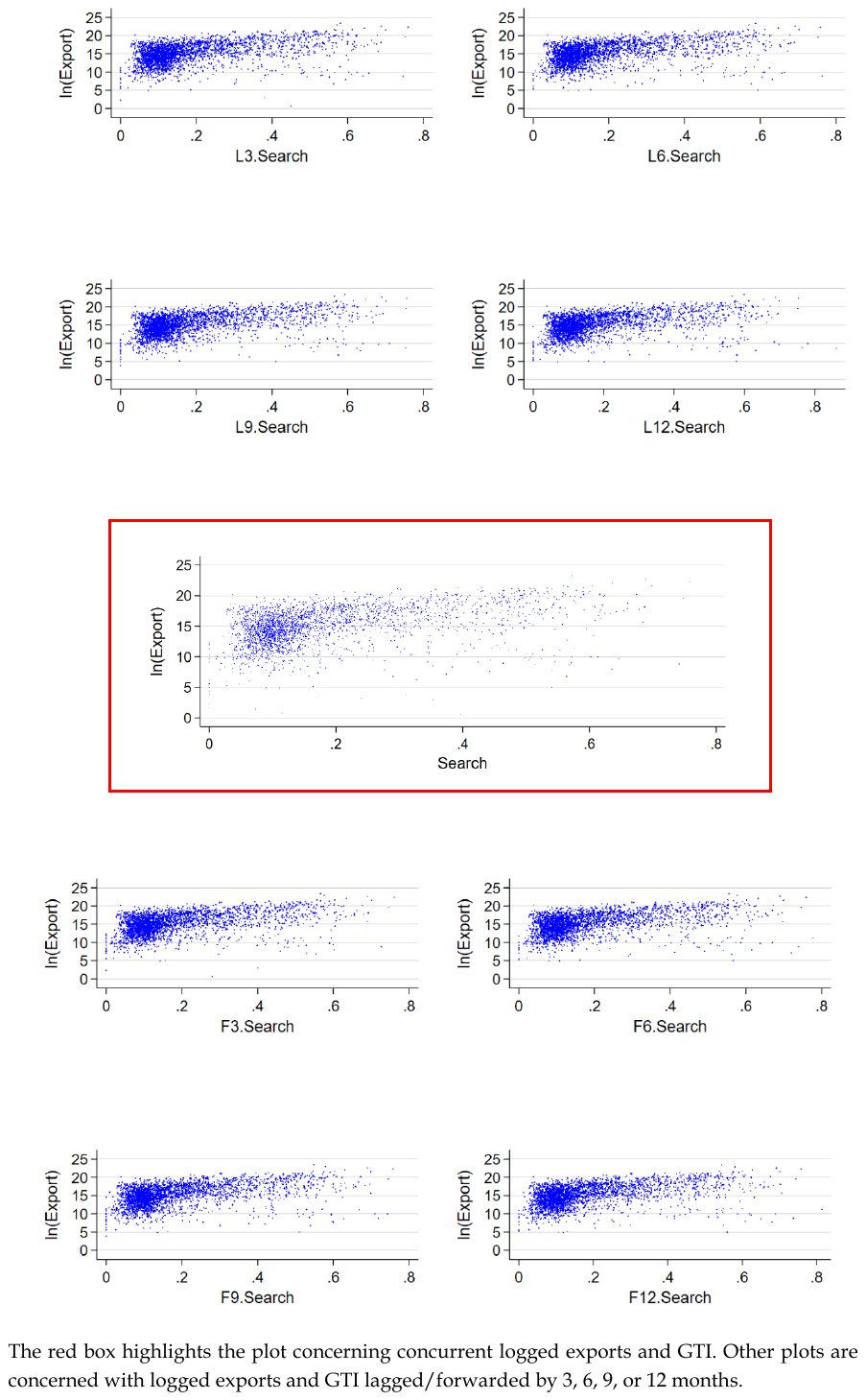}
\vspace{-40pt}
\end{figure}

Before we report the regression results, having an overview of the GTI and export patterns helps understand the data. \Cref{tab:correlation} presents the correlation matrix of \( GTI_{jc\tau} \) for $\tau = t, t\pm 3, t \pm 6,  t\pm 9$, and  $t\pm 12$. The correlation coefficients remain stable over time. The correlation for time periods close to each other is not significantly different than the correlation for periods distant from each other. Also, this stability holds regardless of whether the periods are lagged or forwarded. \Cref{fig:correlation} illustrates the positive relationship between $GTI_{jct}$ and logged exports, where the plot for \( \tau = t \) is highlighted with a red box for emphasis. The figure shows no noticeable variation in the relationship between \( \ln X_{jct} \) and \( GTI_{jc\tau} \) across different \( \tau \) values.  The lack of variation in the relationship across \( \tau \) values is in line with the stable correlation coefficients shown in \Cref{tab:correlation}. Overall, the GTI data are stably paced, viewed either alone or along with the export data.

\begin{table}[h!]
%\vspace{30pt}
\caption{Correlation Matrix} \label{tab:correlation}
\vspace{-40pt}
\hspace*{-1.5cm}\includegraphics[scale=1, page=1]{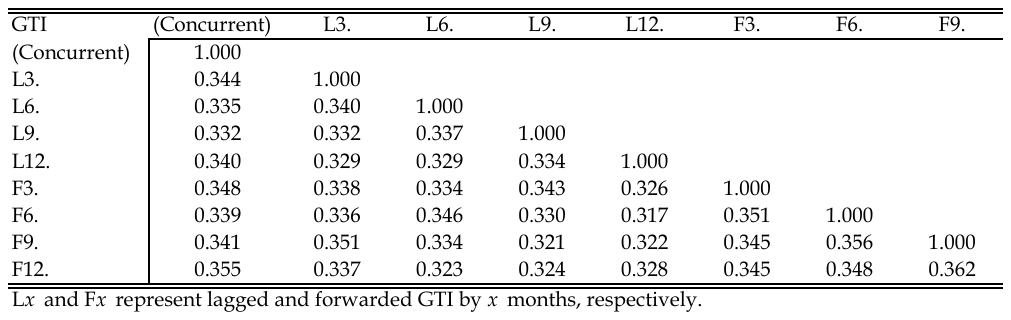}
\vspace{-590pt}
\end{table}

The stably paced GTI, when used as regressors of specification \eqref{eq:reg}, demonstrates explanatory powers distinct from each other. The regression results, reported in \Cref{tab:baseline}, show that only searches conducted in the months leading up to trade are significantly associated with export volume. This positive association suggests that foreign interest in Chinese provinces increases before trade occurs. The GTIs included in the regression span from three to twelve months before and after the trading month. However, incorporating GTIs too close in time can lead to collinearity, reducing statistical significance. In column 2, where $GTI_{jct-3}$, $GTI_{jct}$, and $GTI_{jct+3}$ are included together, $GTI_{jct-3}$ and $GTI_{jct}$ show marginal significance. As the time gap between $GTI_{jct-\tau}$ and $GTI_{jct+\tau}$ increases (i.e., as $|\tau|$ grows), collinearity weakens, leading to higher statistical significance. Additionally, GTIs from distant preceding months show weaker correlations with export volume, a trend observable in the coefficient patterns across columns 2 to 5. In column 6, where all GTIs are included in the regression, only those prior to the trading month remain statistically significant.

\begin{table}[h!]
\vspace{20pt}
\caption{Google Searches and Export Volume: Baseline Results} \label{tab:baseline}
\vspace{-60pt}
\hspace*{-1cm}\includegraphics[scale=1, page=1]{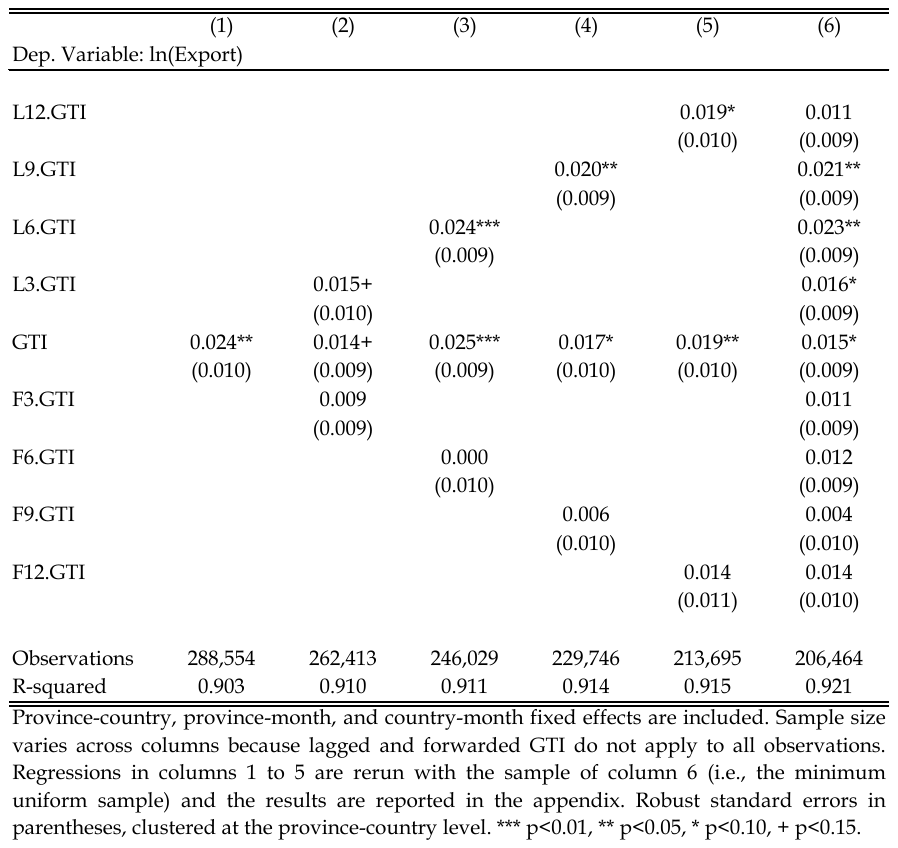}
\vspace{-320pt}
\end{table}
 
% follow the literature. about business cycle data. world bank. etc indicators? 
% time series issue, unit root, etc, discuss

\paragraph{Minimum uniform samples.} Empirical studies in international trade typically use annual origin country-to-destination country data. Our study, which replaces origin countries with origin provinces and annual intervals with monthly intervals, has a much sparser sample space than most trade studies. The sample size in \Cref{tab:baseline} changes slightly across columns because observations missing specific \( GTI_{jc\tau} \) values are excluded from regressions with those \( GTI_{jc\tau} \) terms. Longer lags and forwards, such as \( \tau = t \pm 6 \) compared to \( \tau = t \pm 3 \), tend to result in a greater loss of sample size, since province-country pairs at the start and end of the sample period have fewer \( GTI_{jc\tau} \) values.\footnote{The sample for \( \tau = t \pm 6 \) is not a subset of the sample for \( \tau = t \pm 3 \). Province-country pairs that trade at \( \tau = t \pm 6 \) may not trade at \( \tau = t \pm 3 \), and vice versa. The discrepancy is not necessarily due to differences in trading frequency. Pairs trading every three months would have observations for both \( \tau = t \pm 3 \) and \( \tau = t \pm 6 \). However, pairs trading at irregular intervals may have observations for either \( \tau = t \pm 3 \) or \( \tau = t \pm 6 \) but not both. } Samples that can sustain the complete set of lags and forwards, such as the sample used in column 6 of \Cref{tab:baseline}, are defined as \textit{minimum uniform samples}. They can be used for regressions with an incomplete set of lags and forwards. We use the column 6 sample to rerun the specifications of columns 1--5. The results, reported in \Cref{tab:baseline_rob}, show the same patterns in the magnitude and significance of coefficients as those in \Cref{tab:baseline}. This consistency suggests that the observed patterns in \Cref{tab:baseline} are not attributable to sample differences across columns. Henceforth, we rerun regressions with minimum uniform samples whenever applicable as robustness checks.\footnote{Minimum uniform samples are not our preferred sample setting. First, they have the smallest sample size. Considering the sparsity of monthly trade data, preserving as many observations as possible is essential. Second, minimum uniform samples tend to select province-country pairs that trade frequently over months within a year. Therefore, we rerun regressions with minimum uniform samples solely for robustness checks.}

\paragraph{Margins.} 

Which margin, extensive or intensive, is more sensitive to attention? In international trade, the extensive margin refers to the number of products, such that trade growth at the extensive margin takes the form of additional products being traded and thus extra product lines appearing in customs records. By contrast, the intensive margin denotes the trade volume of given traded products, and thus trade growth at the intensive margin means additional volume of currently existing product lines in customs records. 

We apply the formulas proposed by \citet{HK05} to the province-country-month level, constructing the following extensive and intensive margins of provincial exports:
\begin{equation}
EM_{jct}=\frac{\sum_{i \in I_{jct}}X_{irct}}{\sum_{i \in I}X_{irct}}, \hspace{1cm} IM_{jct}=\frac{\sum_{i \in I_{jct}}X_{ijct}}{\sum_{i \in I_{jct}}X_{irct}},\label{eq:emim}
\end{equation}
where $r$ denotes a reference province other than province $j$, and $i$ denotes a product. By design, province $r$ exports the full set of products $I$ to country $c$ in month $t$, namely $I \equiv \{ \forall i: X_{irct}>0\}$. Intuitively, $EM_{jct}$ isolates the extensive margin variation of $X_{jct}$ by excluding the export volume variation contributed by country $j$ itself---specifically, replacing it with the export volume of reference province $r$. $IM_{jct}$ isolates the intensive margin variation of $X_{jct}$ by counting only the products exported from province $j$ to country $c$ in month $t$ and deflating the export volume by reference province $r$.
\begin{table}[h!]
%\vspace{30pt}
\caption{Extensive and Intensive Margins} \label{tab:margins}
\vspace{-10pt}
\hspace*{1cm}\includegraphics[scale=.7, page=1]{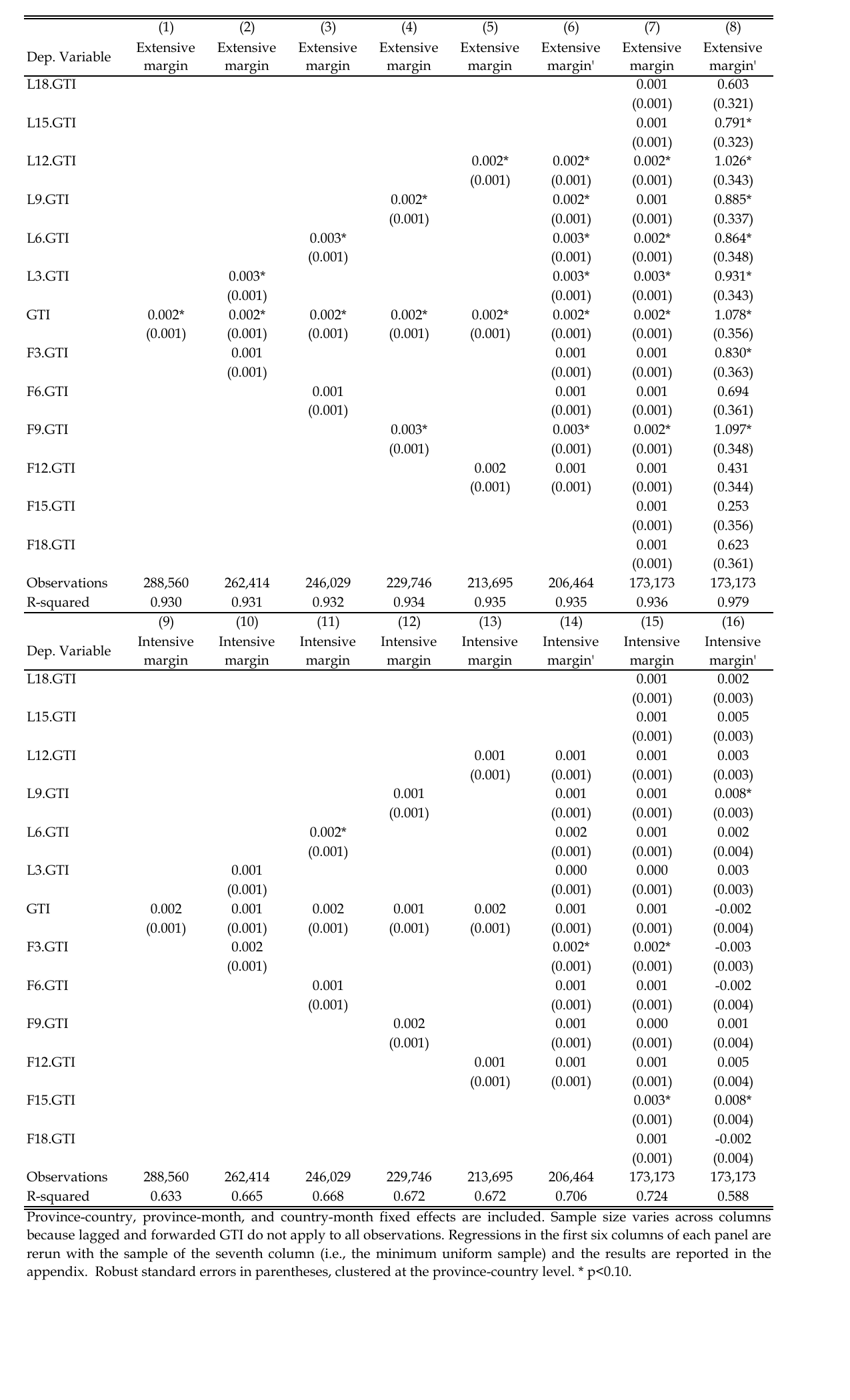}
%\vspace{-320pt}
\end{table}

The merit of the above $EM_{jct}$ and $IM_{jct}$ is that they are measures relative to a reference country such that they are not driven by the width of product categories. Product categories in international trade statistics build on  the Harmonized System (HS) classifications administered by the World Customs Organization. A given new product tends to be classified as an extensive (intensive) margin increase if it belongs to a fine (broad) category. The use of reference countries in $EM_{jct}$ and $IM_{jct}$ addresses such arbitrariness caused by HS classifications. Nonetheless, counting product categories in exports remains the most straightforward approach to defining the extensive margin. Thus, we construct an alternative extensive margin measure based on simple counting and an alternative intensive margin measure based on the value per product category:
\begin{equation}
EM_{jct}'=\frac{N_{jct}}{N_{ct}}, \hspace{1cm} IM_{jct}'=\frac{X_{jct}/N_{jct}}{X_{ct}/N_{ct}}, \label{eq:emim2}
\end{equation}
where $N_{jct}$ is number of HS4 product lines exported by province $j$ to country $c$ in month $t$,  $N_{ct}$ is the number of HS4 product lines exported from Mainland China to country $c$ in month $t$, and $X_{ct}$ is the total export volume from Mainland China to country $c$ in month $t$.\footnote{The product of $EM_{jct}$ and $IM_{jct}$ is the ratio of province $j$'s to reference province $r$'s exports to country $c$ in month $t$, or
\begin{equation*}
EM_{jct} IM_{jct} = \frac{\sum_{i \in I_{jct}}X_{ijct}}{\sum_{i \in I}X_{ikct}}=\frac{X_{jct}}{X_{rct}}.
\end{equation*}
In comparison, the product of $EM_{jct}'$ and $IM_{jct}'$ equals 
\begin{equation*}
EM_{jct}' IM_{jct}' = \frac{X_{jct}}{X_{ct}}, \label{eq:marginproduct2}
\end{equation*}
which is the share of province $j$'s exports in China's total exports to country $c$ in month $t$.} 
 
We apply specification \eqref{eq:reg} to dependent variables $EM_{jct}$, $IM_{jct}$, $EM_{jct}'$ and $IM_{jct}'$.  To maximize the set $I$, we use the rest of China from each province $j$'s perspective as its reference province $r$. The results are reported in \Cref{tab:margins}. The extensive margin of exports turn out positively associated with concurrent and preceding GTI, while the intensive margin shows little such association. That is, foreign attention, as measured by GTI, is more closely linked to a greater variety of trade rather than to a higher volume of trade within a given variety.\footnote{We rerun the regressions in columns 1 to 5 with the sample of column 6 (i.e., the minimum uniform sample) and reach the same findings. The detailed results are reported in \Cref{tab:margins_rob}. }

\paragraph{Imports.} 

We also apply the specifications of \Cref{tab:baseline} and \Cref{tab:margins} to the import data of Chinese provinces, the results from which are reported in \Cref{tab:importside}.\footnote{Descriptive statistics for the import data are reported in \Cref{tab:summarystats_imports}.} Interestingly, the correlation between GTI's various lags/forwards and export volume does not hold on the import side. As shown, there barely exists a correlation between GTI and imports. Just as for the export side, we rerun the regressions in column 1 to 5 with the sample of column 6 (the minimum uniform sample) and the finding, as reported in \Cref{tab:importside_rob1}, remains the same. Considering that China's trade partners are not an identical set, we single out China's import-source countries that are also its export-destination countries and rerun the regressions in \Cref{tab:importside} with imports from those countries. The finding, as reported in \Cref{tab:importside_rob2}, remains the same.\footnote{The minimum-uniform-sample robustness check for columns 1--5 of \Cref{tab:importside_rob2} is reported in \Cref{tab:importside_rob3}.} 

% the provinces: elaborate. 

\begin{landscape}
\begin{table}[h!]
%\vspace{30pt}
\caption{Google Searches and Import Volume} \label{tab:importside}
%\vspace{-20pt}
\hspace*{0cm}\includegraphics[scale=.9, page=1]{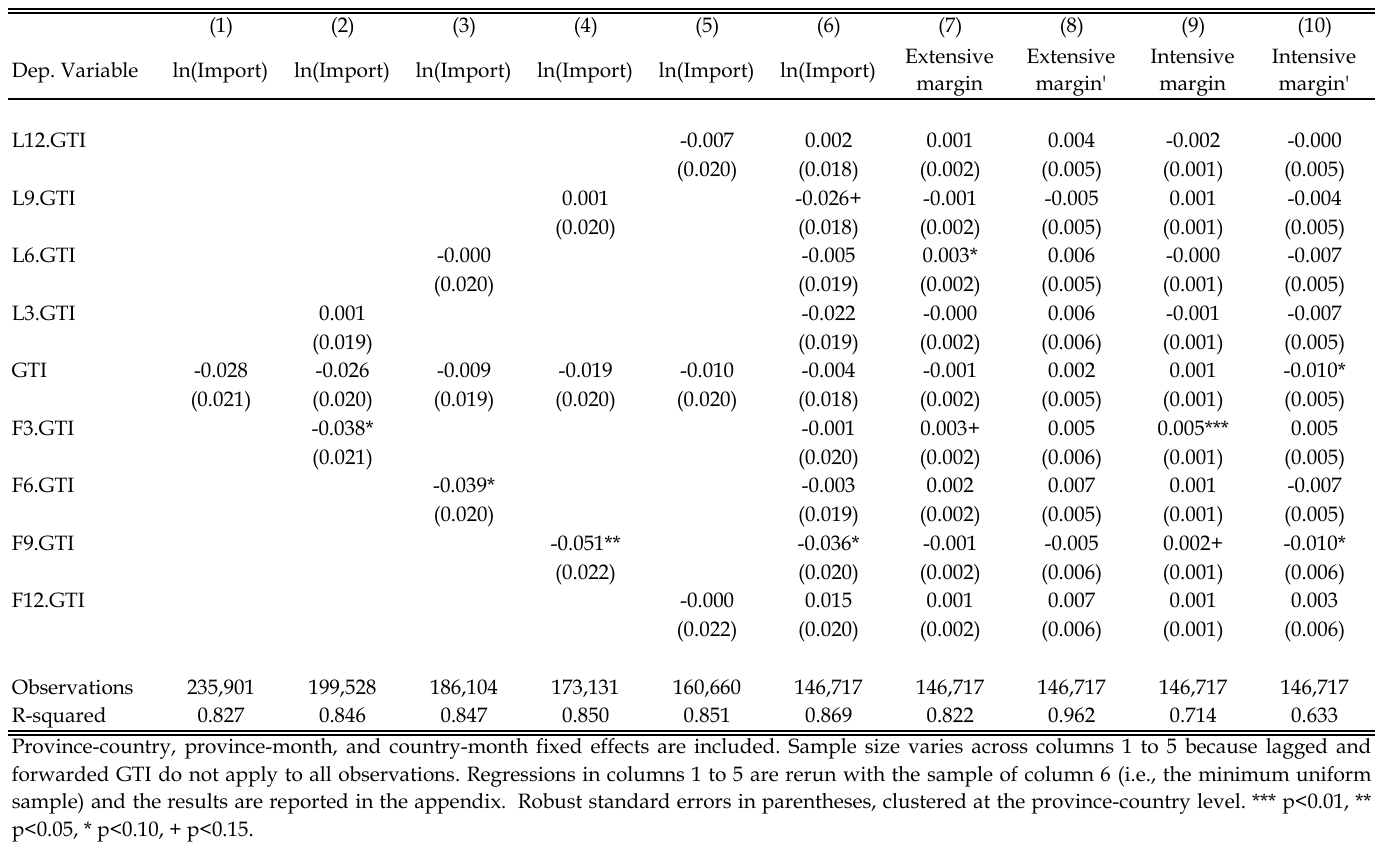}
%\vspace{-320pt}
\end{table}
\end{landscape}

A natural explanation for the absence of correlation is that the relevance of attention to China's trade, as measured by GTI, stems from the products made by Chinese provinces but not from the foreign products purchased by those provinces. In other words, foreign importers who are interested in Chinese provincial export supplies search those provinces more, while foreign exporters who are interested in Chinese provincial import demands do not particularly search for those provinces. In essence, Google is primarily an import-side tool.\footnote{Chinese importers may search foreign places for foreign products. Since Google is blocked in China, there exist no reliable data recording those potential searches. VPN-based Google searches are too selected to represent Chinese importers' Google searches.}

\paragraph{Heterogeneity.} 

Now we examine whether the above GTI-exports relationship varies by product section. The HS classifications used in global trade consist of 21 sections at the top level of their hierarchy. We use HS Section 3 (animal and vegetable oil) as the reference group. By interacting section dummies with GTI and controlling for province-country-HS4, province-month, and country-month fixed effects, we estimate the differing association between GTI and exports across HS sections. The results are reported in \Cref{tab:HSsec}. The interaction terms were constructed with concurrent GTI, while interaction terms constructed with lagged GTI give similar results (available upon request). The association is stronger for products involving natural resources such as minerals, leather, vegetables, and pulp and paper. 

\begin{table}[h!]
%\vspace{30pt}
\caption{By HS Section} \label{tab:HSsec}
\vspace{-10pt}
\hspace*{1cm}\includegraphics[scale=.85, page=1]{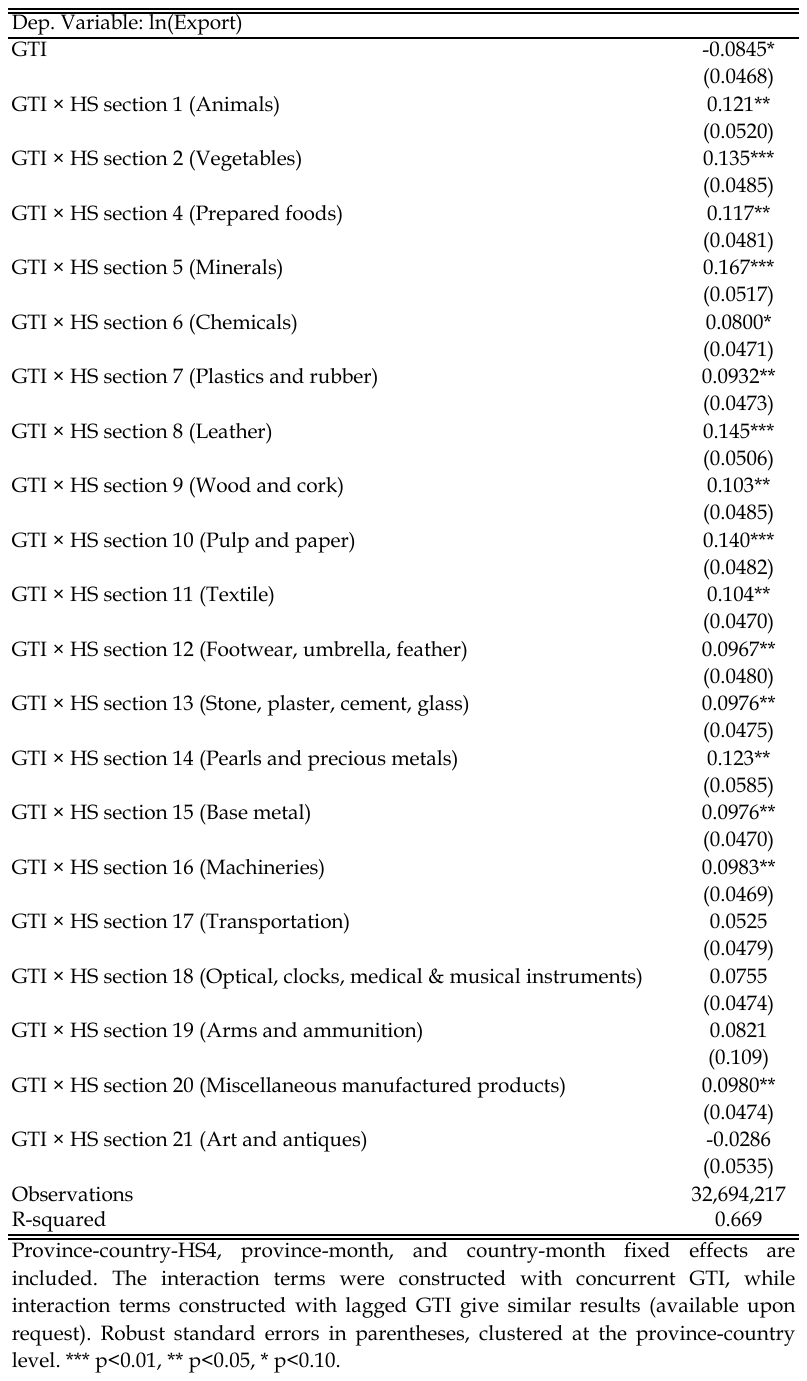}
\vspace{-120pt}
\end{table}

\begin{table}[h!]
%\vspace{30pt}
\caption{Country and Product Characteristics} \label{tab:interact}
\vspace{-20pt}
\hspace*{0cm}\includegraphics[scale=.9, page=1]{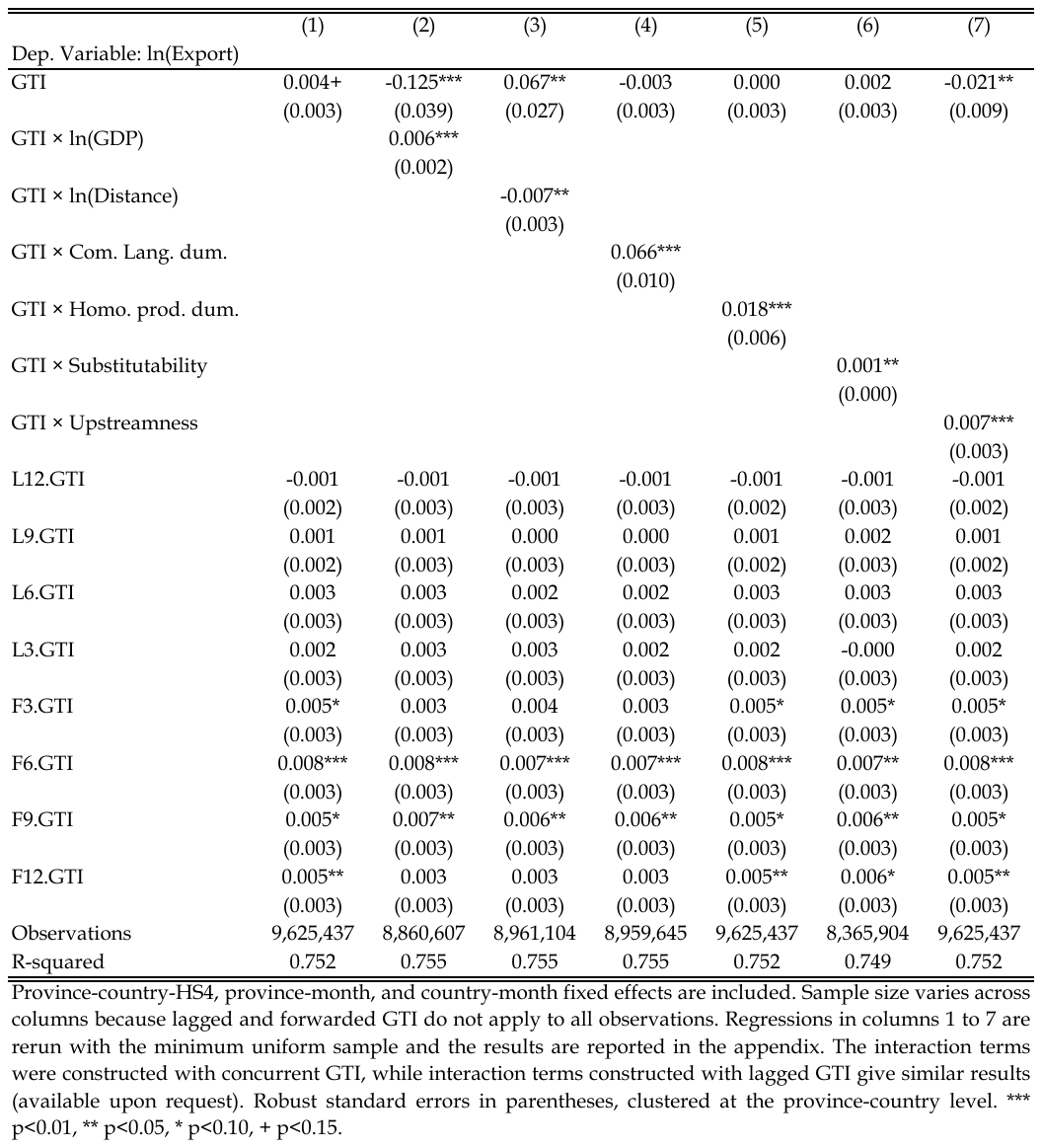}
\vspace{-180pt}
\end{table}

In \Cref{tab:interact}, we interact GTI with country and product characteristics. The country characteristics considered here are sourced from the gravity model literature, including destination-country GDP, distance, and whether the destination country speaks a common language with China (i.e., Chinese).\footnote{\citet{HMR10}, who used these variables to estimate gravity models, made the data publicly available at \url{https://www.cepii.fr/CEPII/en/bdd_modele/bdd_modele.asp}.} Product characteristics considered here include whether the products in focus are homogeneous, their substitutability, and their upstreamness in the production process. Whether products are homogeneous or not is determined according to \citet{Rauch99}. That is, products that are traded on organized exchanges or have reference prices in commodity trade journals are defined as homogeneous since their brands and manufacturer identities are relatively unimportant in determining their features. Substitutability, measured by the elasticity of substitution, was estimated by \citet{BW06} with a CES demand structure. Perfect complements (respectively, substitutes) have an elasticity of zero (respectively, positive infinity). The upstreamness of products was estimated by \citet{CMY21} using Chinese data, measuring the location of a product in the production process. For instance, rubber, which can be used either as a final product or an intermediate input, is considered more upstream than apparel, which is rarely used as an intermediate input.\footnote{The original data on homogeneity classifications and the elasticities were constructed across SITC codes while the upstreamness data were reported across Chinese industry codes. We converted them to HS4 codes.} 

As shown, the correlation between GTI and exports is higher for larger and closer destination countries and countries that speak the same language. Also, the correlation is stronger for products that are relatively homogeneous, substitutable, and upstream in the production process.\footnote{We rerun the regressions in columns 1--7 with the sample of column 8 (minimum uniform sample) as a robustness check. The results, as reported in \Cref{tab:interact_rob}, are consistent with those in \Cref{tab:interact}.} 

Taking \Cref{tab:baseline} through \Cref{tab:interact} together, a clear explanation emerges for why and how Chinese exports are linked to foreign attention. Provinces that receive more attention from foreign importers are more likely to export local products to those countries. Among the products made locally, those that are relatively homogeneous, substitutable, and upstream in the production process are still more likely to benefit from the additional foreign attention. The resulting increases in exports primarily contribute to the extensive margin rather than to the intensive margin, because web searches are more likely to result in the formation of new trade relations between exporters and importers rather than increasing the trade volume within existing exporter-importer relations. In contrast, existing trade relations, whether ongoing or discontinued, seldom involve web searches.

\subsection{The search elasticity of trade \label{sec:searchelas}}

We now estimate the (Google) search elasticity of trade. The gravity model literature proposes the following method to estimate elasticities in international trade \citep{HMR10,HM14}. Assume that the characteristics of exporters and importers are multiplicative, then taking the ratios relative to a reference exporter (province) and a reference importer (country) removes those characteristics and isolates the elasticity of interest:\footnote{Consider $X_{jct}=M_{jt}^{\alpha}M_{ct}^{\zeta}GTI_{jct}^{\delta}$, where $M_{jt}$ and $M_{ct}$ represent province-month and country-month characteristics, respectively. In the expression $\frac{X_{jct}/X_{jdt}}{X_{kct}/X_{kdt}}$, $M_{jt}^{\alpha}$ is canceled out by  $M_{kt}^{\alpha}$, and $M_{ct}^{\zeta}$ is canceled out by $M_{dt}^{\zeta}$.}
\begin{equation}
\ln \frac{X_{jct}/X_{jdt}}{X_{kct}/X_{kdt}}=\delta \ln \frac{GTI_{jct}/GTI_{jdt}}{GTI_{kct}/GTI_{kdt}} + \frac{\epsilon_{jct}/\epsilon_{jdt}}{\epsilon_{kct}/\epsilon_{kdt}} ,\label{eq:ratio_ratio}
\end{equation}
where $k$ is a reference province (relative to province $j$) and $d$ is a reference country (relative to country $c$). Unlike the $\beta$'s in specification \eqref{eq:reg}, the estimated elasticity $\delta$ is not conditional on province or country characteristics that are controlled for. 

The reference provinces and countries in specification \eqref{eq:ratio_ratio}  need to be representative. Not all Chinese provinces export to a given country, nor do all countries import from a given Chinese province. We use the rest of Chinese provinces (from province $j$'s perspective) as the reference province $k$, and the rest of the importing countries (from country $c$'s perspective) as the importing country $d$. 

\begin{table}[h!]
%\vspace{30pt}
\caption{Google-search Elasticity} \label{tab:elasticity}
\vspace{-40pt}
\hspace*{0.5cm}\includegraphics[scale=1, page=1]{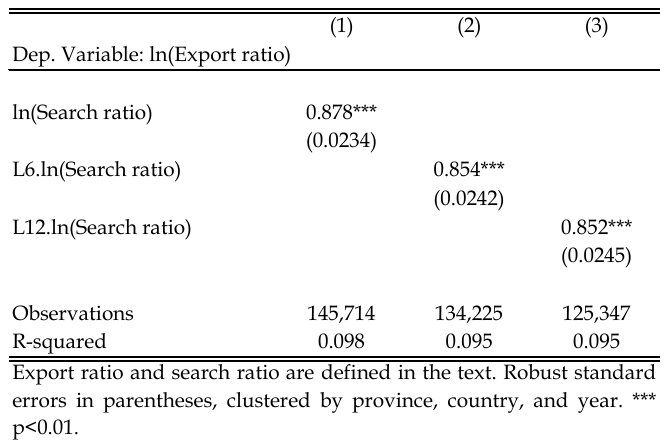}
\vspace{-500pt}
\end{table}

The estimation results for the elasticity $\delta$ are reported in \Cref{tab:elasticity}, where the export ratio and search ratio refer to the dependent and independent variables in specification \eqref{eq:ratio_ratio}. Concurrent GTI is used in column 1, where the $\delta$-estimate is around 0.88. That is, when the relative GTI rises by one percent, the export volume is expected to rise by 0.88 percent. GTI is lagged by 6 and 12 months in columns 2 and 3, respectively, and the $\delta$-estimate drops to around 0.85 when the length of lags increases. The elasticity can be used to predict export volume along the cross-sectional dimension. For instance, if the trade statistics of a regional economy are suspected to be inaccurate or unreliable, the web search data can help validate the trade statistics with trade statistics from regional economies that are known to be reliable. 

\section{Extensions \label{sec:extensions}}

The previous section explored the relationship between GTI and trade volume (and its two margins). With additional data, methods, or contexts, the GTI data can be utilized to investigate various dimensions of international trade. In this section, we illustrate such potentials of GTI by extending the previous analysis into four distinct applications.

\subsection{Extension A: Trading prices \label{sec:extensionA}}

The first extension is concerned with the prices in international trade. The  publicly available customs data from China are at the province-country-month-HS8 level, enabling us to compute price metrics at the province-country-month-HS4 level. The first group of price metrics we examine include mean, minimum, and maximum prices. We also compute the coefficient of variation (i.e., the standard deviation divided by the mean) of prices. \Cref{tab:prices} reports how the correlation between these price metrics and GTI varies by product characteristics (the same ones as in \Cref{tab:interact}). As shown, the coefficients of interest are generally insignificant, except for a decrease in price dispersion for homogeneous products relative to non-homogeneous products as GTI increases.

The lack of price sensitivity to GTI aligns with our earlier finding that the link between online attention and trade operates mainly through the extensive margin. In other words, online attention facilitates the formation of new trade relations, while its relevance to trade within established relations remains minimal.

\begin{landscape}
\begin{table}[h!]
%\vspace{30pt}
\caption{Trading Prices} \label{tab:prices}
%\vspace{-20pt}
\hspace*{1cm}\includegraphics[scale=1, page=1]{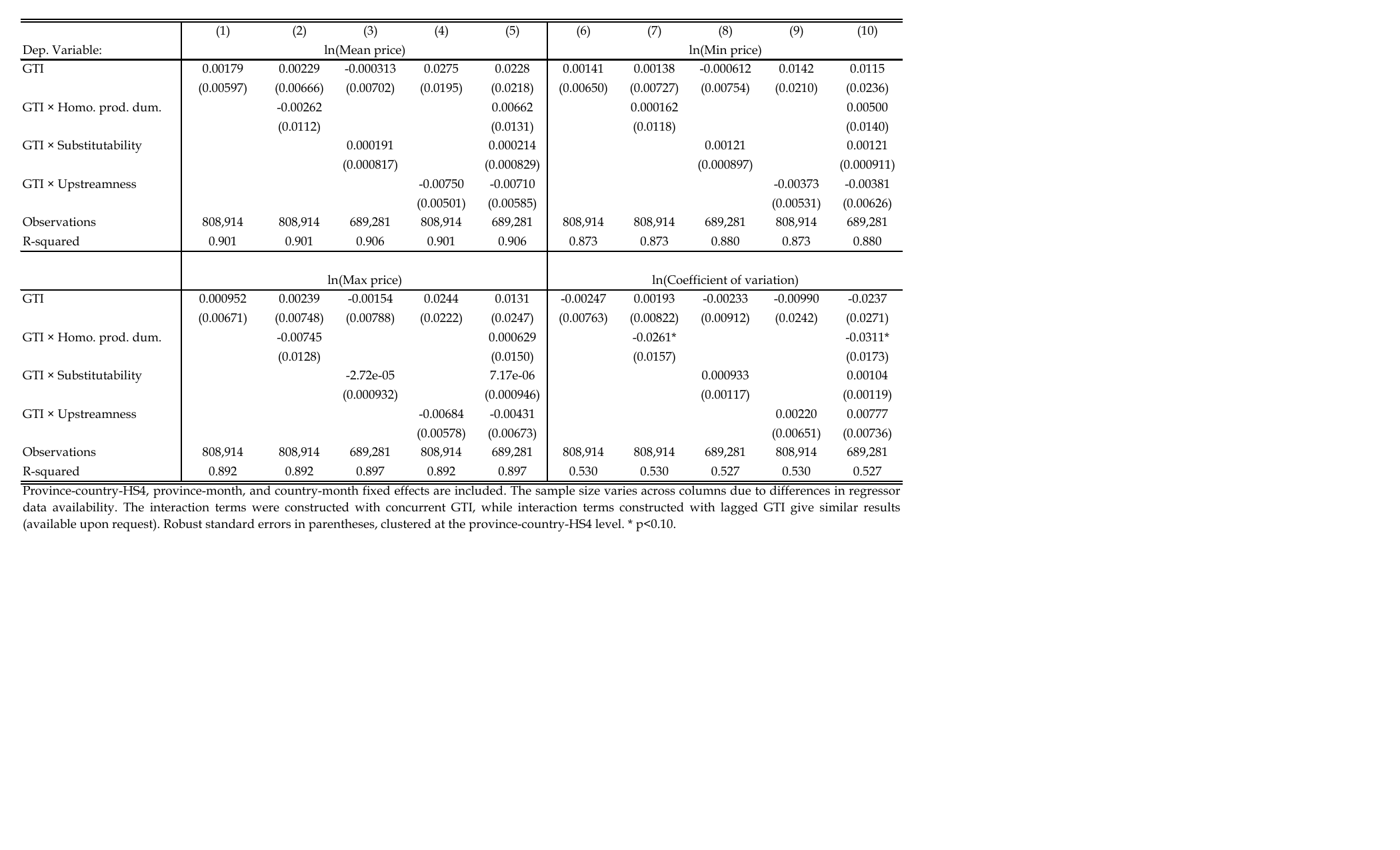}
\vspace{-120pt}
\end{table}
\end{landscape}

\subsection{Extension B: Export forecasting \label{sec:extensionB}}

Returning to the province-country-month level data, the second extension of our study examines the effectiveness of GTI predictors in forecasting exports. We limit the number of GTI predictors to one, as increasing the number of predictors raises both data requirements and the risk of overfitting. In \Cref{tab:forecast}, Model 1 reproduces the last column of \Cref{tab:baseline}, serving the purpose of comparison. Model 2 keeps only the concurrent GTI. Model 3 includes concurrent GTI and its polynomials of degree up to 4. Models 4--6 uses only one lagged GTI, which can be 1-month, 6-month, or 12-month. In Models 7-9, we consider a hypothetical scenario where we use forwarded GTI to predict concurrent exports. 

By comparing across columns, we reach four findings. First, the GTI from recent lagged months exhibits strong predictive power. Second, concurrent GTI provides moderate predictive strength---comparable to the 6-month lag but weaker than the 3-month lag---while its polynomial terms add little value. Third, forward-looking terms, which are unavailable in real forecasting scenarios, contribute minimally to forecast accuracy. Fourth, overall, the 1-month lag GTI demonstrates the highest predictive power. These findings indicate that past search activity helps forecast present trade patterns, whereas future search data, even hypothetically available at the time of forecasting, do not. Concurrent GTI underperforms compared to the 1-month lag GTI due to its potential forward components. These findings reinforce our earlier baseline findings, showing that Google searches from previous months have a significant statistical association with exports, whereas those from later months do not.

\begin{landscape}
\begin{table}[h!]
\vspace{-20pt}
\caption{Trade Forecasting} \label{tab:forecast}
%\vspace{-20pt}
\hspace*{1cm}\includegraphics[scale=.9, page=1]{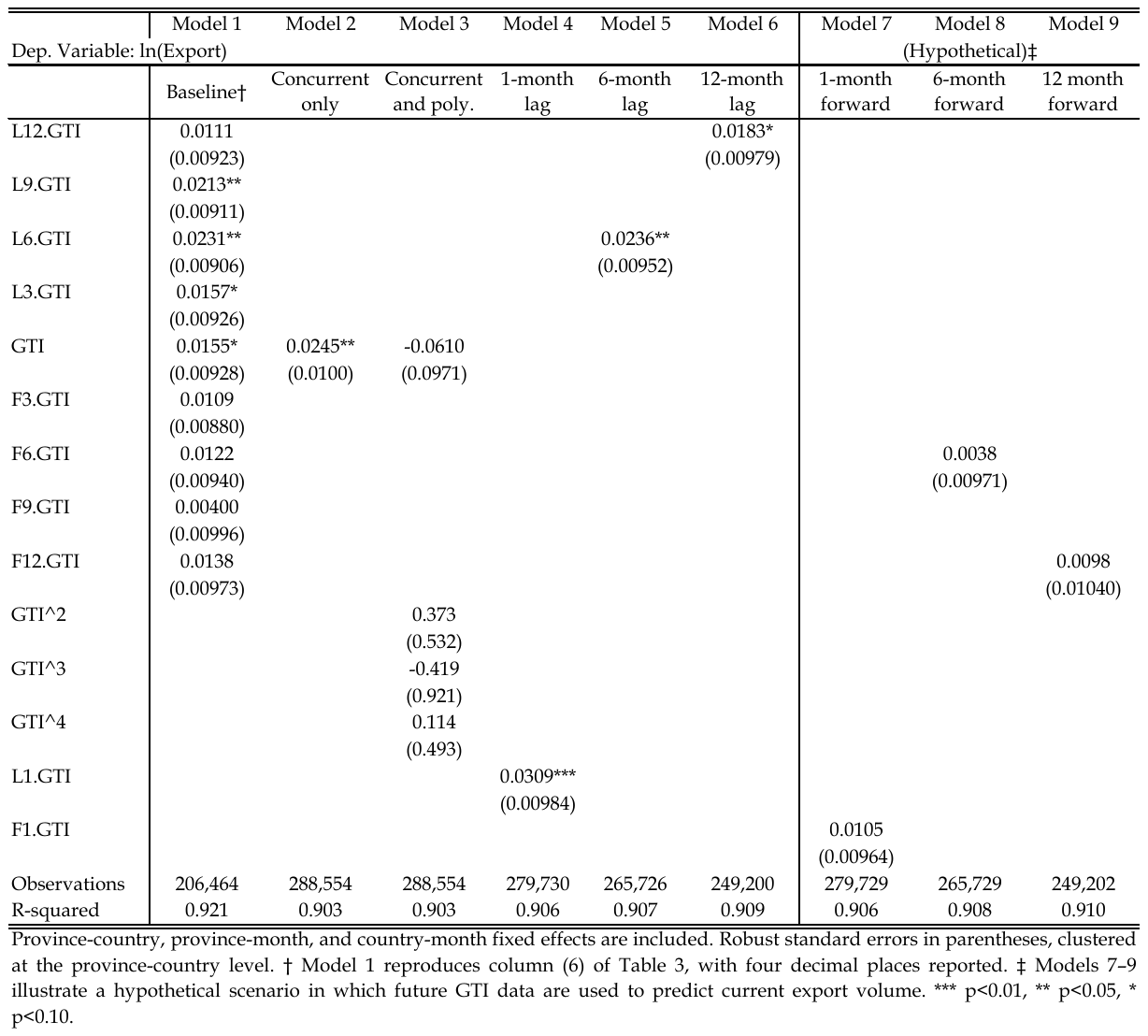}
\vspace{-320pt}
\end{table}
\end{landscape}

\subsection{Extension C: Trade in a pandemic \label{sec:extensionC}}

\begin{figure}[h!]
%\vspace{30pt}
\caption{Search Frequency for the Keyword \textit{Hubei}} \label{fig:hubei}
%\vspace{-10pt}
\hspace*{-.5cm}\includegraphics[scale=.85, page=6]{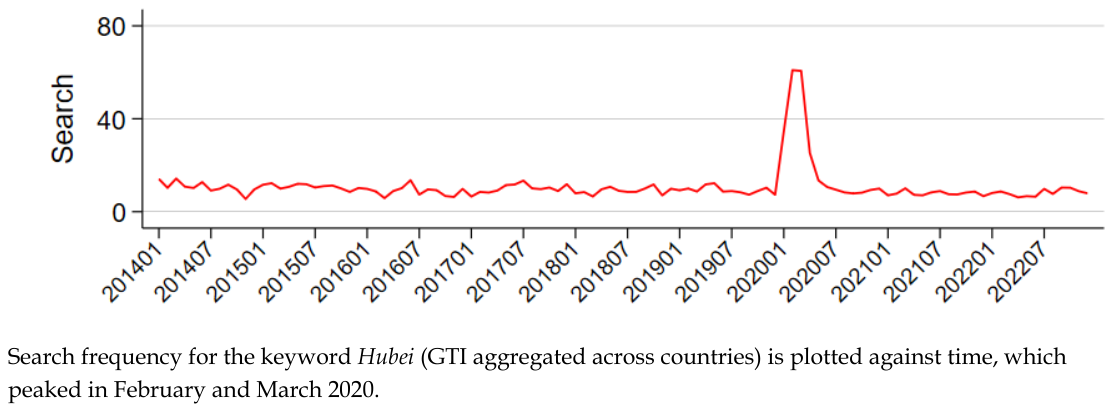}
\vspace{-540pt}
\end{figure}

The COVID pandemic originated in China's Hubei province in late 2019 and quickly spread globally, causing widespread illness, economic disruption, and lockdowns in the years that followed. \Cref{fig:hubei} illustrates the elevated Google search frequency for the keyword \textit{Hubei}, reaching its peak in February and March 2020 as the virus rapidly spread from the region to the rest of the country and the world. In response to the pandemic, governments around the globe implemented measures like social distancing, travel restrictions, and vaccination campaigns to control the spread. The internet became a more important information channel during the pandemic than in normal times. China, as the largest exporter in the world, was also the country with the strictest testing and quarantine measures in the world. Those measures intermittently disrupted international traveling from and to China during the pandemic.\footnote{During the COVID pandemic, all travelers entering China, regardless of nationality, were subject to mandatory quarantine. Initially set at 14 days, this requirement was later extended to up to 21 days. For a detailed discussion of China's quarantine policy, see \citet{Ba23}.} The third extension of our study asks: Did COVID make web searches more influential to China's exports? 

We specified the following regression to estimate the role of the COVID pandemic in the relationship between Google searches and exports:
\begin{equation}
\ln X_{jct} = \beta_\tau GTI_{jc\tau} + \gamma_\tau GTI_{jc\tau} \times  \boldsymbol{1}[t \text{ in COVID time}] + \lambda_{jc} + \lambda_{jt} + \lambda_{ct} + \epsilon_{jct}, \label{eq:reg_covid}
\end{equation} 
where $\tau$ ranges between $t-12$ and $t+12$. COVID-month indicator $\boldsymbol{1}[t \text{ in COVID time}]$ equals 1 for months starting from January 2020. The Wuhan lockdown represented the beginning of China's travel restrictions, which lasted until the end of our sample (October 2022). The coefficient of interest, $\gamma_{\tau}$, captures how the association between Google searches and concurrent exports changed when the pandemic began.

\begin{table}[h!]
%\vspace{30pt}
\caption{Google Searches and Export Volume during the COVID Pandemic} \label{tab:covid}
\vspace{-40pt}
\hspace*{-0cm}\includegraphics[scale=.9, page=1]{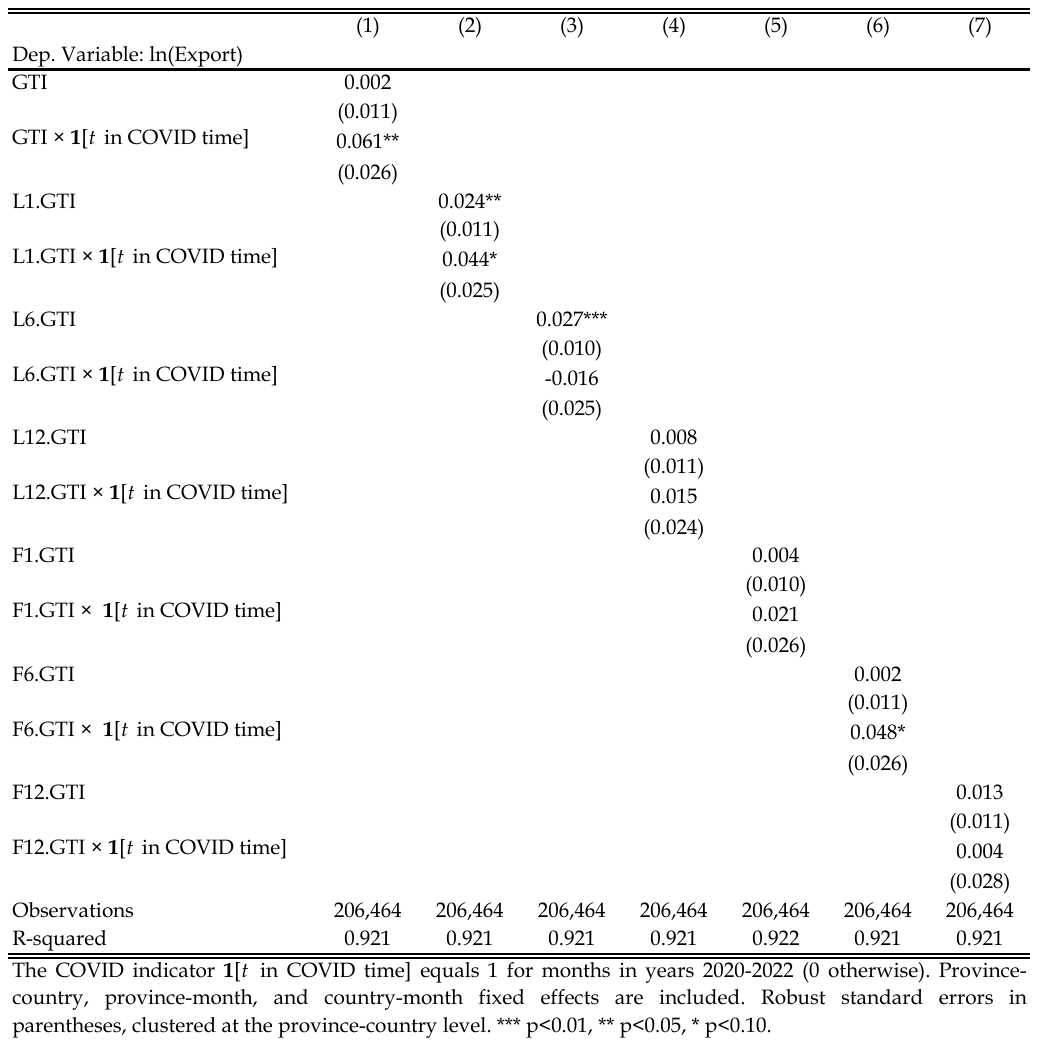}
\vspace{-220pt}
\end{table}

The results from specification \eqref{eq:reg_covid} are reported in \Cref{tab:covid}. The explanatory power of prior Google searches for exports turns out to be strengthened by the pandemic. The additional strength in prediction applies only to the Google searches conducted in preceding months. Serving as placebo checks, the same regression using Google searches from succeeding months (of equivalent lengths) shows no similar effect.  

\begin{figure}[h!]
%\vspace{30pt}
\caption{Google Searches and Exports during the COVID Pandemic: by Month} \label{fig:covid}
%\vspace{-20pt}
\hspace*{0.5cm}\includegraphics[scale=.7, page=7]{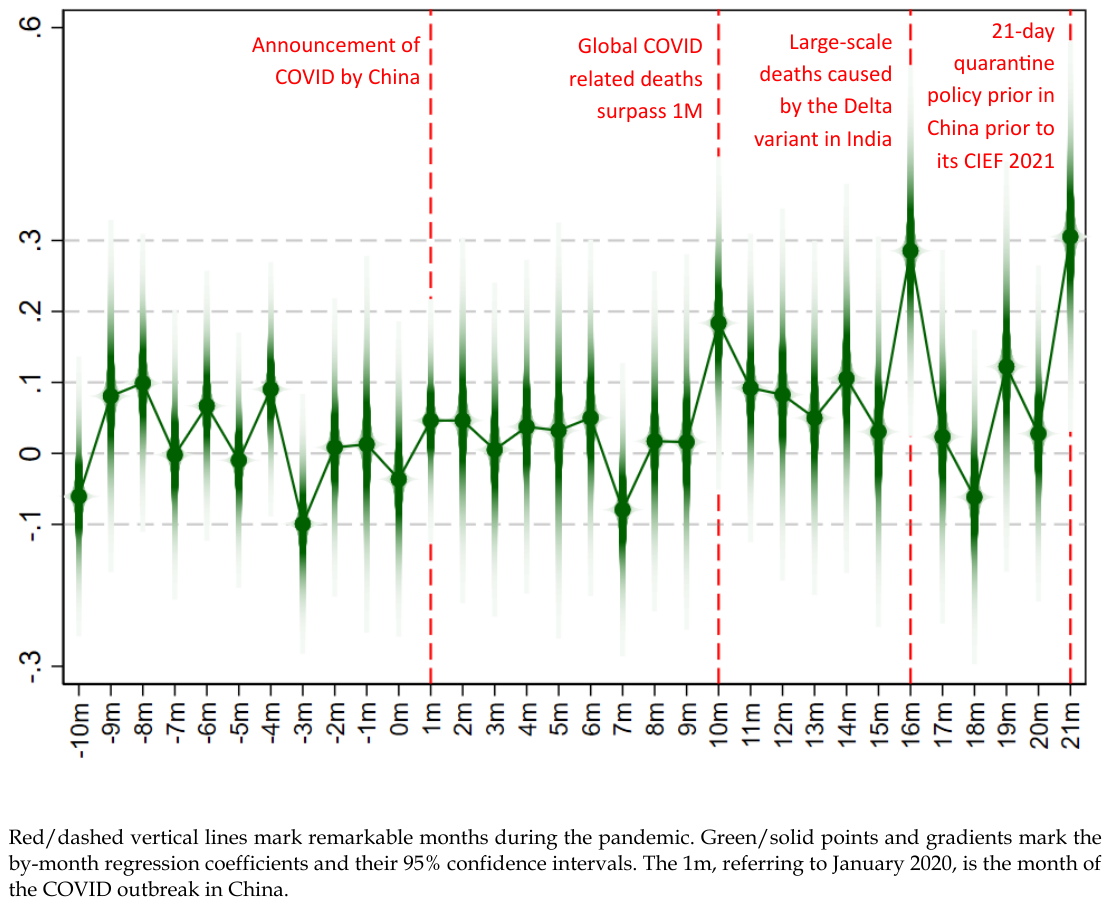}
\vspace{-260pt}
\end{figure}

COVID-month indicator $\boldsymbol{1}[t \text{ in COVID time}]$ in regression \eqref{eq:reg_covid} divides our sample period into pre-COVID and COVID periods. We also experiment with calendar month indicator $\boldsymbol{1}[t]$ instead of $\boldsymbol{1}[t \text{ in COVID time}]$ to see if the GTI-export relationship varies over time.\footnote{Uninteracted month indicators are equivalent to month fixed effects, which have been included in all regressions.} The coefficients of interactions are reported in \Cref{fig:covid} along with their 95\% confidence intervals. The four months marked in the figure represented four salient moments during the global pandemic: (i) the official announcement of the pandemic and control measures by the Chinese government (January 2020), (ii) global COVID-related deaths surpassed one million (October 2020), (iii) large-scale deaths occurred in India due to the Delta variant (April 2021), and (iv) China released its 21-day quarantine requirement before its China Import and Export Fair 2021 (September 2021). Both the severity of the pandemic and the tightening of preventive polices magnified the importance of web searches in information collection. As expected, we find a stronger association between Google searches and export volume in the months of those four events.

\subsection{Extension D: Seasonality in trade \label{sec:extensionD}}
This extension examines the seasonality of international trade, focusing on the holiday peak in Chinese exports from November to January driven by elevated foreign demand. Foreign attention, including searches by individual consumers, importers, retailers, and producers who import inputs, likely amplifies this seasonal pattern. Outside the holiday season, internet searches play a lesser role as trade is dominated by steady demand and long-term contracts. We hypothesize that the relationship between provincial exports and Google searches is most pronounced during the holiday season.

\begin{table}[h!]
\caption{Google Searches and Seasonality in Provincial Exports}
\vspace{-60pt}
\label{tab:seasonality}
\vspace{-0pt}
\hspace*{.9cm}\includegraphics[scale=1, page=1]{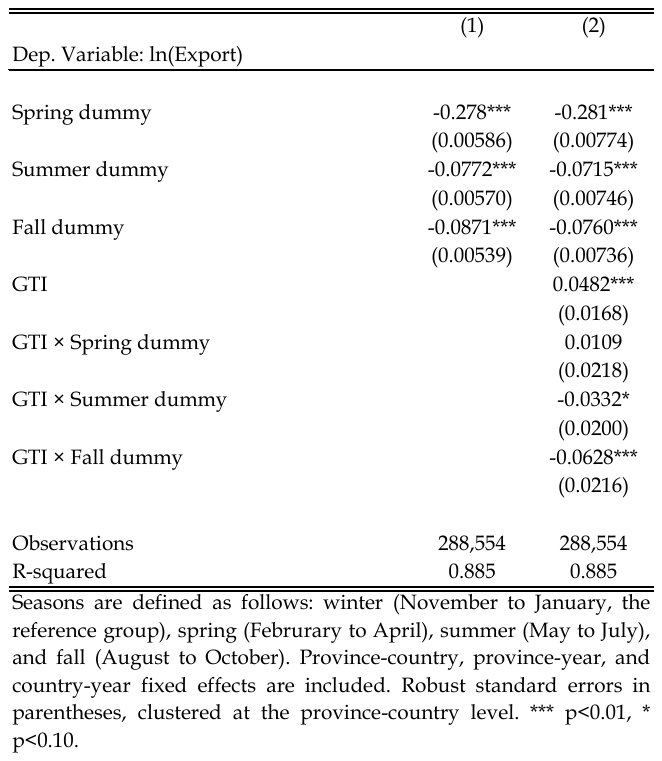}
\vspace{-360pt}
\end{table}

Column 1 of \Cref{tab:seasonality} confirms the seasonality of provincial exports. The regression includes three seasonal dummies---Spring (February to April), Summer (May to July), and Fall (August to October)---with Winter (November to January), the holiday season, as the reference group. Otherwise, the specification mirrors column 1 of \Cref{tab:baseline}, except that province-month and destination-month fixed effects are replaced with province-year and destination-year fixed effects. The negative coefficients on all three seasonal dummies underscore the elevated import demand during the holiday season. Column 2 incorporates interactions between seasonal dummies and GTI, revealing a weaker association between GTI and exports in summer and fall. In summary, while Google searches for China do not exhibit seasonality, their importance in the Chinese export business varies across seasons.\footnote{The lack of seasonality in the GTI data is demonstrated in \Cref{fig:provsVSchina} through \Cref{fig:examplecountries} and supported by the unit-root test in \Cref{tab:unitroot}.} This is consistent with our earlier findings, indicating that foreign attention mainly influences the import side.

\section{Concluding Discussion \label{sec:conclude}}
A universal principle applies to businesses at all levels---local, regional, national, or global: buyers pay attention to a place before purchasing from its sellers. However, measuring attention in international trade remains a complex challenge. We leverage Google search data to track the online attention Chinese provinces receive from abroad. Our findings indicate that provinces export significantly more to countries that have searched for them in the past 12 months, with an estimated elasticity of 0.85--0.88. This online attention premium is particularly evident at the extensive margin of exports, especially for products that are relatively homogeneous, substitutable, and upstream in the production process. The effect is still more pronounced during the COVID pandemic and during holiday seasons compared to other periods.

The interest of foreign importers and the citizens they serve may either predate the searches or arise at the time the searches are conducted. In the first case, Google searches execute existing interest, while in the second, they generate new interest. In both scenarios, Google searches capture the scarce attention that foreign demand allocates across Chinese provinces. If attention were not scarce, search patterns would not correlate with trade, as all provinces would eventually receive searches as needed, making the timing of searches irrelevant to the timing of trade.

Future research could develop a theoretical framework to examine the interplay between trade and web searches. An ideal model would capture the dynamic allocation of attention, distinguishing between exogenous attention shocks and endogenous attention usage to better guide empirical analysis. For instance, foreign buyers may first encounter a location through random web searches, develop an interest in it, and then conduct targeted web searches for sellers or products. This sequence highlights both roles of web searches in trade: executors of existing interest and generators of new interest. While web search data alone cannot differentiate these roles, a theoretical model could provide the necessary structure to do so.

%\newpage
\bibliographystyle{te}
\bibliography{google}

%\subsection{Figures}

%this is figure \ref{fig:fig1}

\begin{appendices}

\appendix

\section*{\center ``Google and China's Trade''  \\ Online Appendix }

\begin{center}
Cui Hu and Ben G. Li \\
\today
\end{center}
\renewcommand{\thesection}{A}
\renewcommand{\thetable}{A}
\renewcommand{\thefigure}{A}
\setcounter{figure}{0} \renewcommand{\thefigure}{A\arabic{figure}}
\setcounter{table}{0} \renewcommand{\thetable}{A\arabic{table}}
\numberwithin{equation}{section}
\setcounter{equation}{0}  % reset counter

\begin{table}[h!]
\caption{Tests for Unit Roots}
\centering
\label{tab:unitroot}
\vspace{-20pt}
\hspace*{-1cm}\includegraphics[scale=.9, page=1]{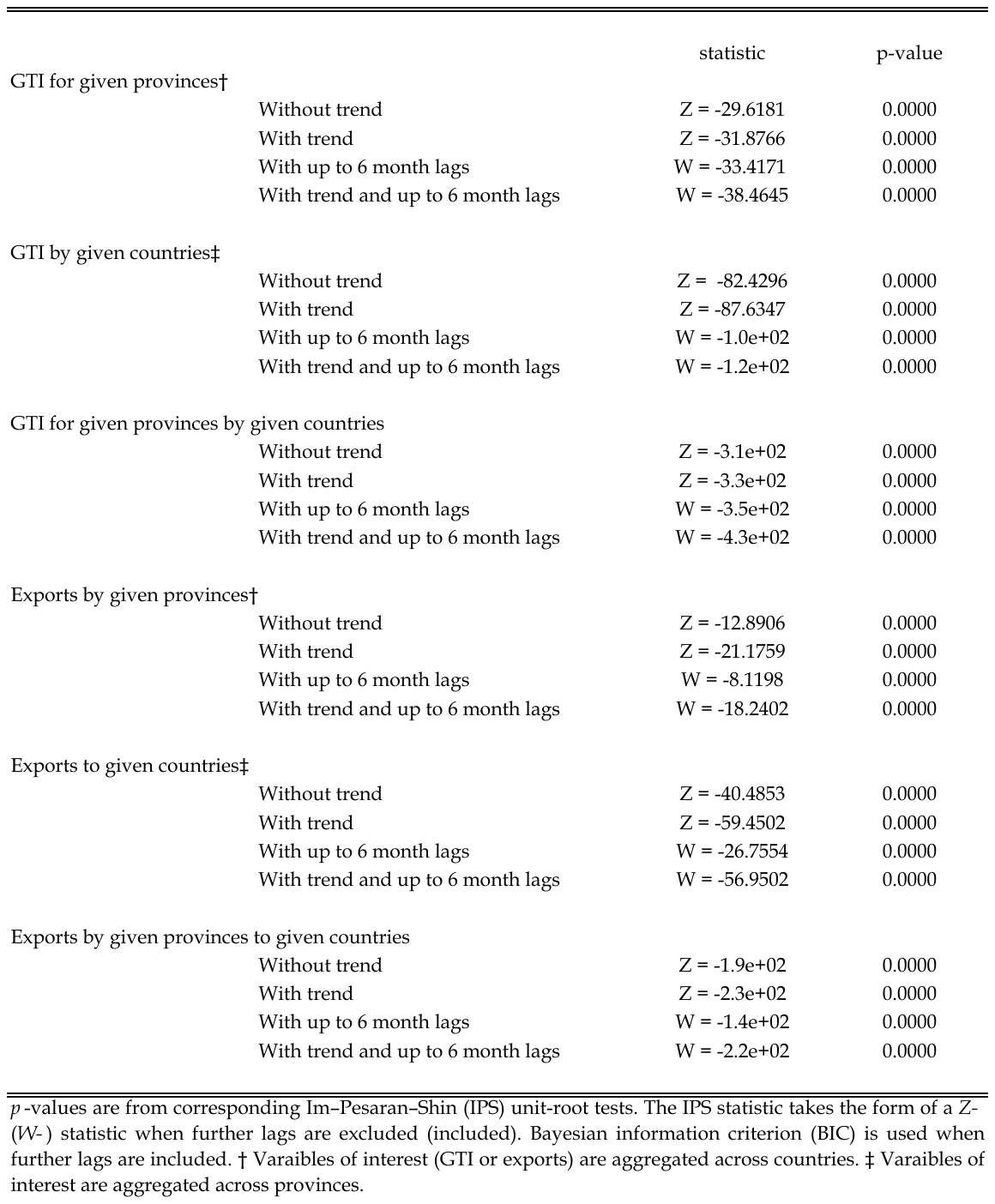}
\end{table}

\begin{table}[h!]
\caption{Baseline Results (First Five Columns with the Minimum Uniform Sample)}
\centering
\label{tab:baseline_rob}
\vspace{-20pt}
\hspace*{-0cm}\includegraphics[scale=1, page=1]{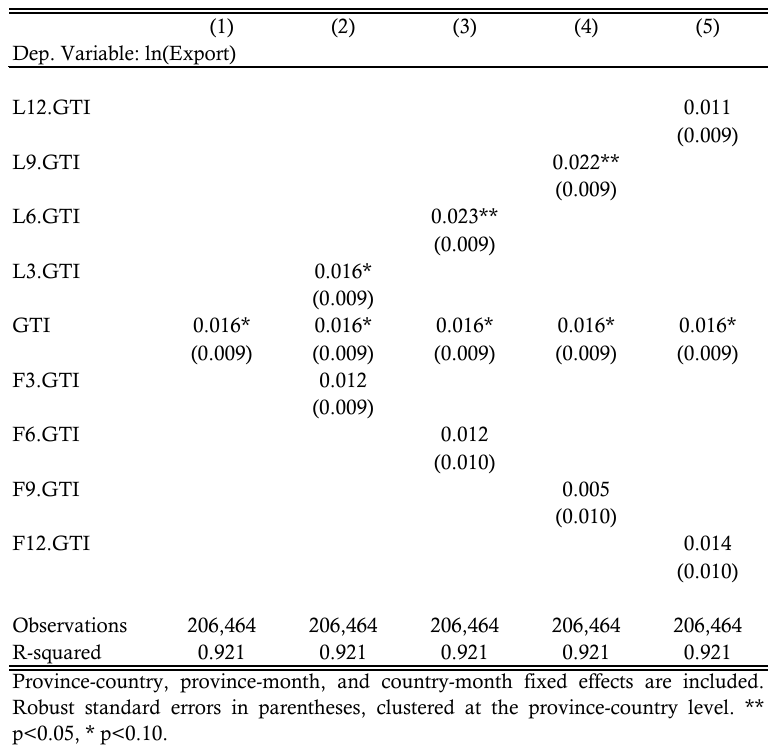}
\end{table}

\begin{table}[h!]
%\vspace{30pt}
\caption{Extensive and Intensive Margins (First Five Columns with the Minimum Uniform Sample)} \label{tab:margins_rob}
\vspace{-20pt}
\hspace*{3cm}\includegraphics[scale=.75, page=1]{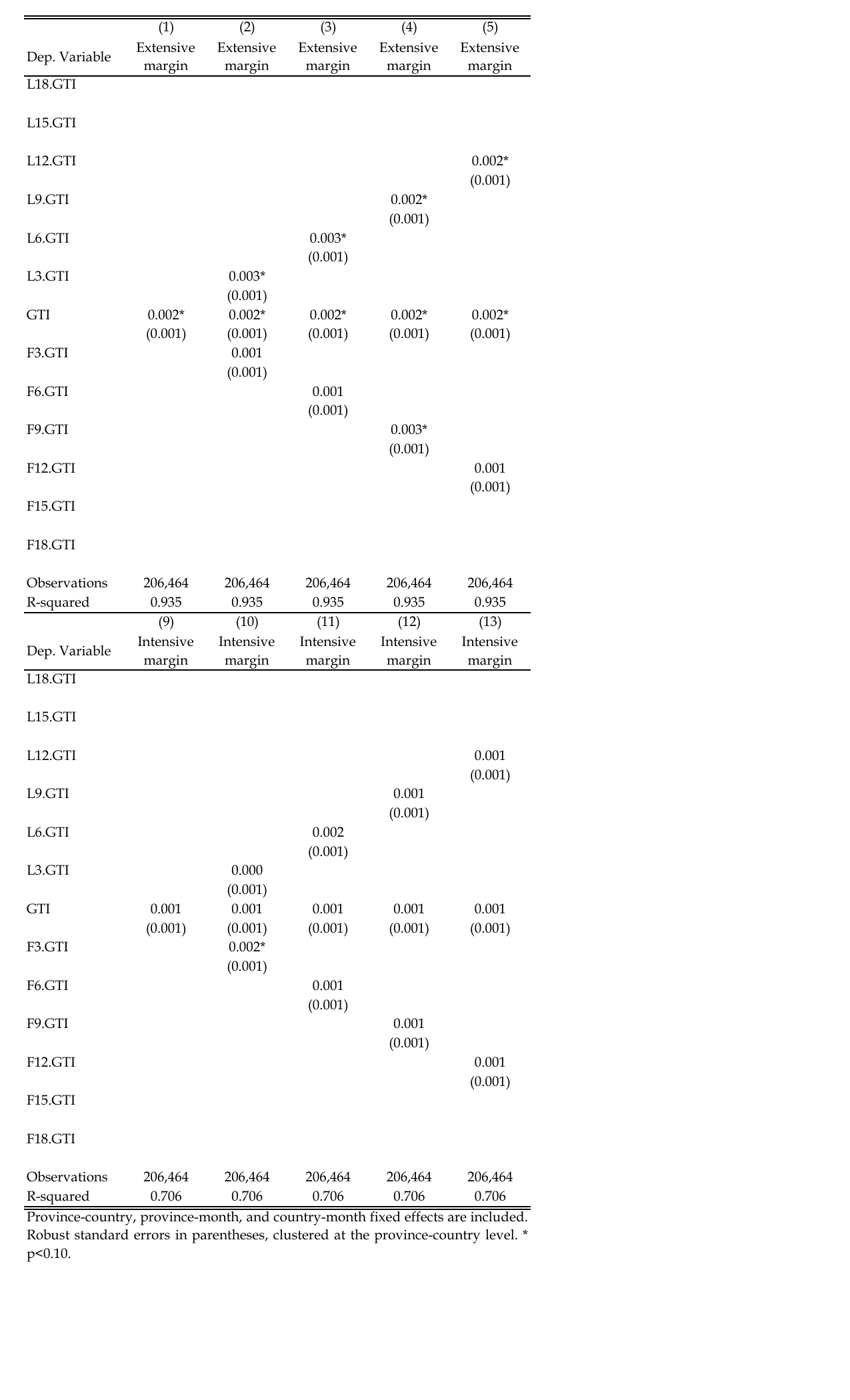}
%\vspace{-320pt}
\end{table}

\begin{table}[h!]
\vspace{20pt}
\caption{Descriptive Statistics for the Imports Data} \label{tab:summarystats_imports}
\vspace{-40pt}
\hspace*{0cm}\includegraphics[scale=1, page=1]{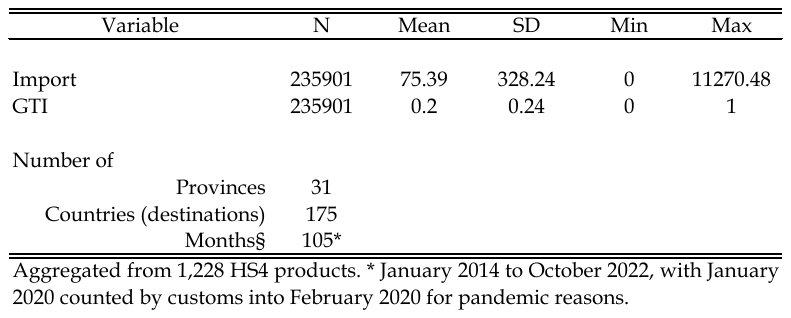}
\vspace{-600pt}
\end{table}

\begin{table}[h!]
\caption{Google Searches and Import Volume (First Five Columns with the Minimum Uniform Sample)}
\centering
\label{tab:importside_rob1}
\vspace{-10pt}
\hspace*{1cm}\includegraphics[scale=1, page=1]{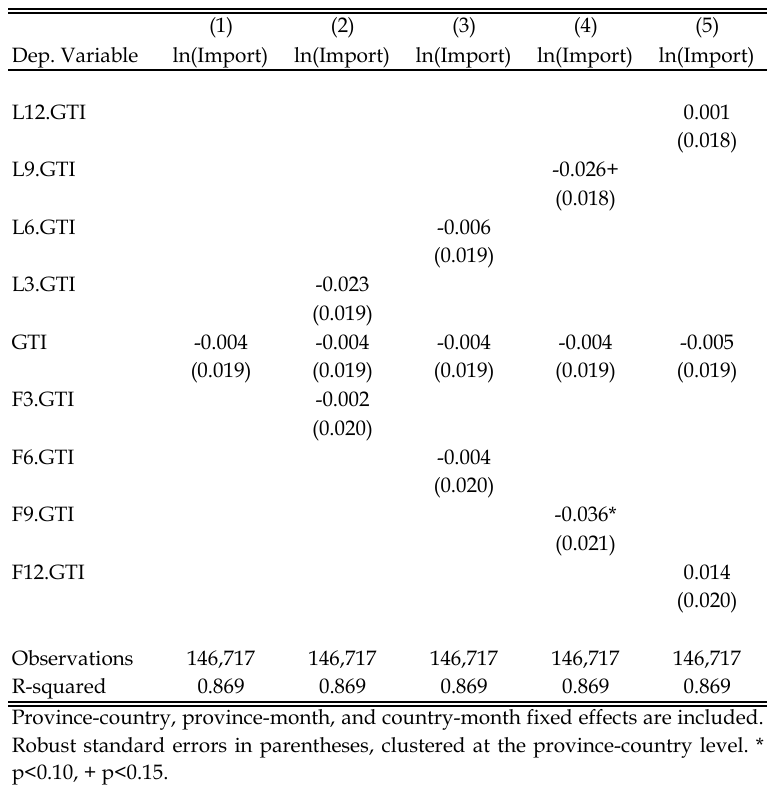}
\end{table}

\begin{landscape}
\begin{table}[h!]
%\vspace{30pt}
\caption{Google Searches and Imports (Import Sources being Export Destinations) } \label{tab:importside_rob2}
%\vspace{-20pt}
\hspace*{0cm}\includegraphics[scale=.9, page=1]{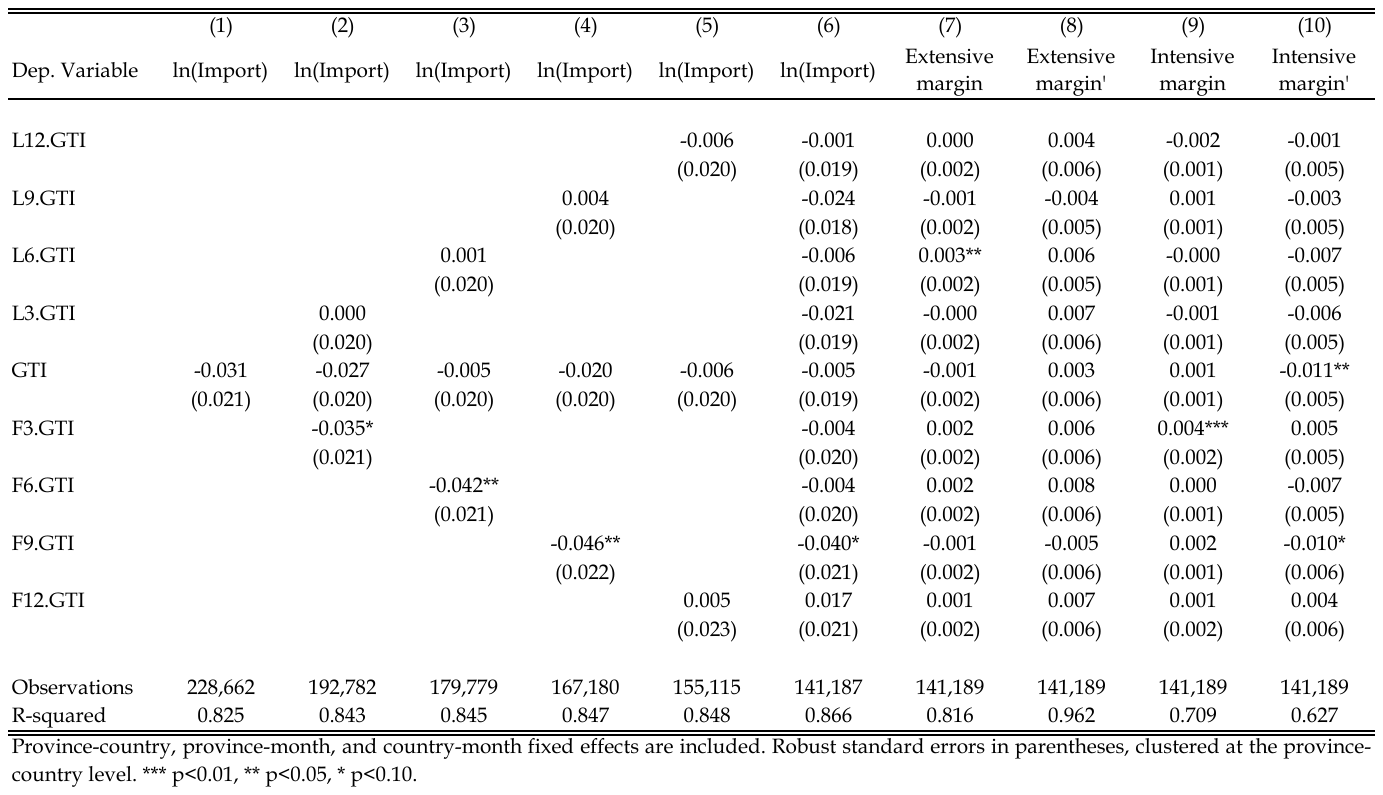}
%\vspace{-320pt}
\end{table}
\end{landscape}

\begin{table}[h!]
\caption{Google Searches and Imports (Import Sources being Export Destinations, First Five Columns with the Minimum Uniform Sample)}
\centering
\label{tab:importside_rob3}
\vspace{-0pt}
\hspace*{1cm}\includegraphics[scale=1, page=1]{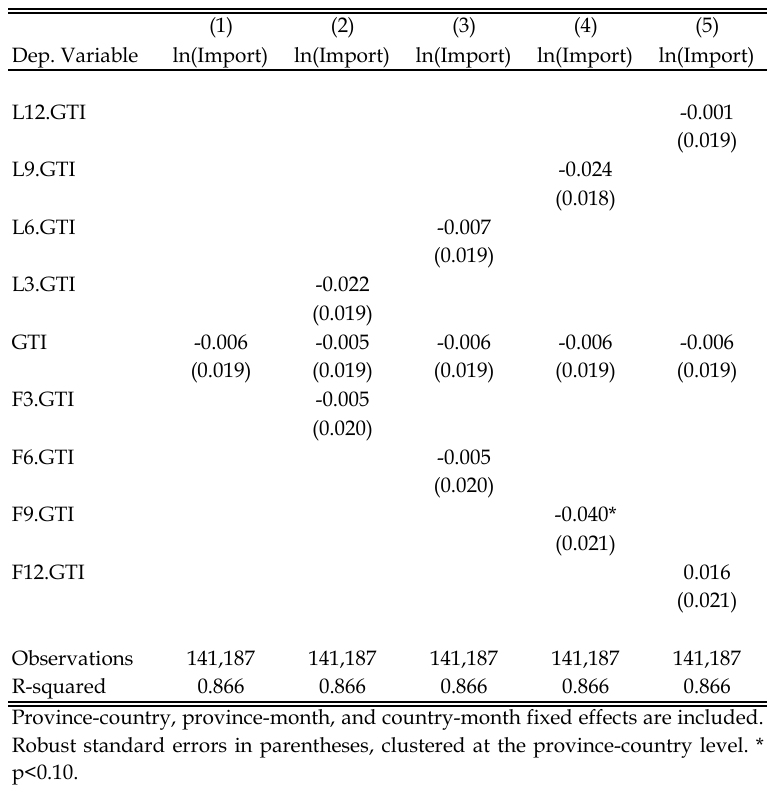}
\end{table}

\begin{table}[h!]
%\vspace{30pt}
\caption{Country and Product Characteristics (All Columns with the Minimum Uniform Sample)} \label{tab:interact_rob}
\vspace{-20pt}
\hspace*{0cm}\includegraphics[scale=.9, page=1]{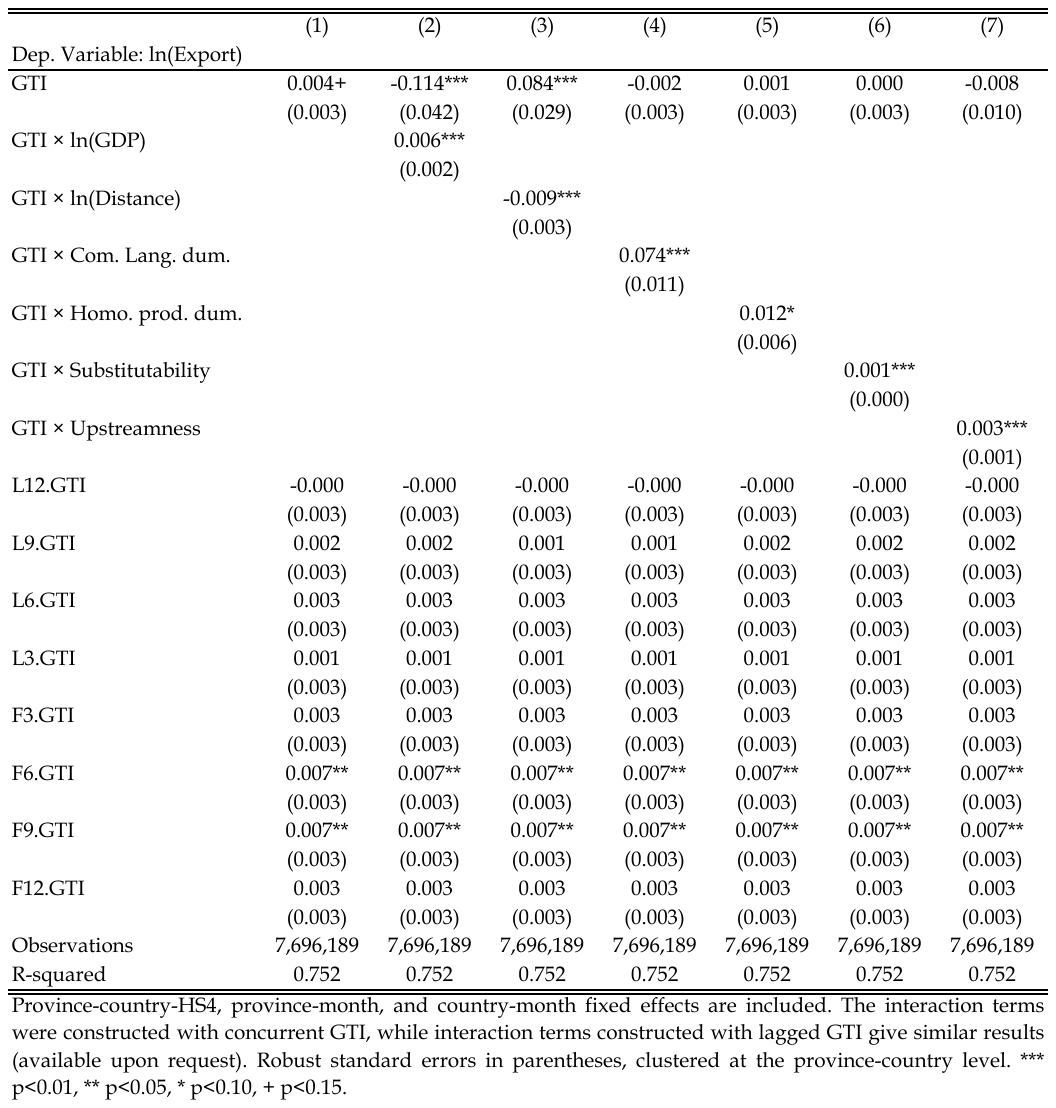}
\vspace{-220pt}
\end{table}

\begin{landscape}
\begin{figure}[h!]
%\vspace{30pt}
\caption{Province-product Combinations Used as Search Keywords } \label{fig:provinceproduct}
%\vspace{-10pt}
\hspace*{.5cm}\includegraphics[scale=.9, page=1]{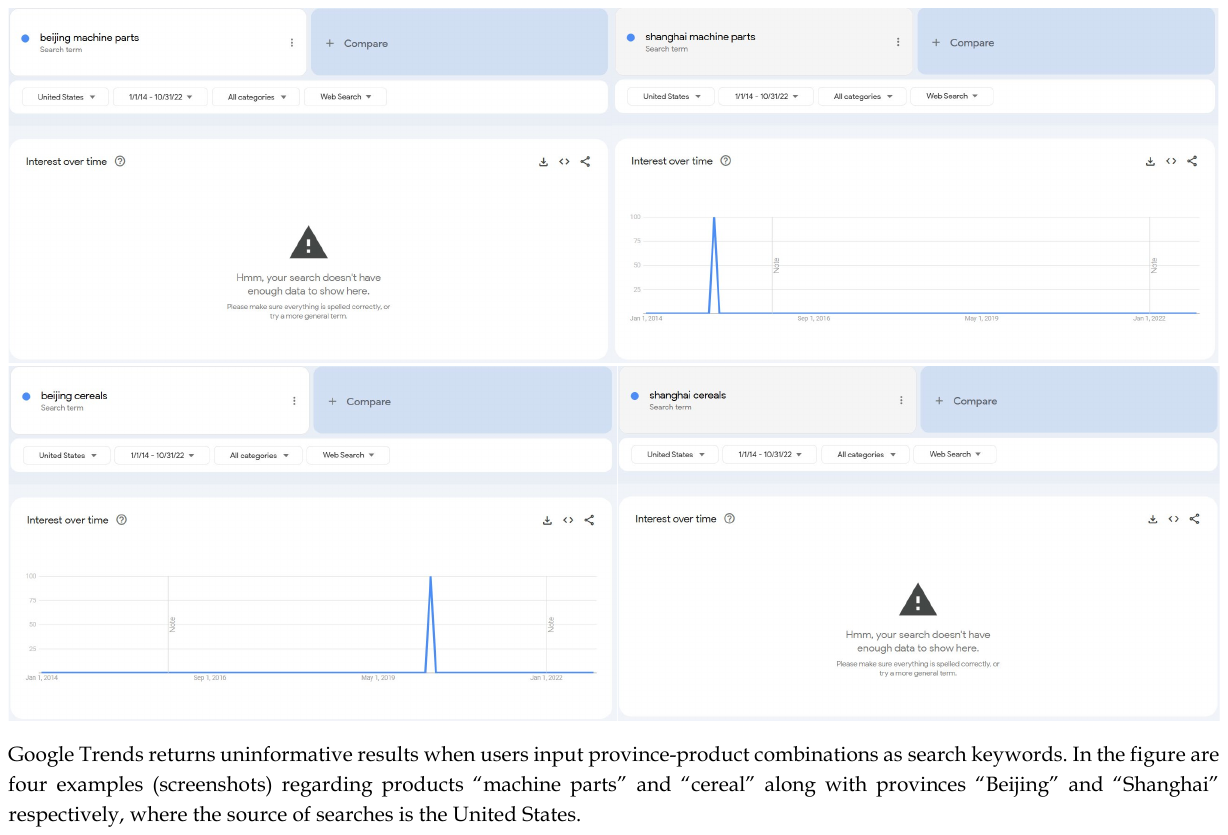}
%\vspace{-420pt}
\end{figure}
\end{landscape}

\end{appendices}

\end{document}